\documentclass[reprint,floatfix,showpacs]{revtex4-1}

\usepackage{graphicx}
\usepackage{dcolumn}
\usepackage{amsmath}
\usepackage{amssymb}
\usepackage{color}

\begin{document}

\title{Propagating Fronts in Fluids with Solutal Feedback}
\author{S. Mukherjee}
\affiliation{Department of Biomedical Engineering and Mechanics, Virginia Tech, Blacksburg, Virginia 24061}
\author{M. R. Paul}
\email{Corresponding author: mrp@vt.edu}
\affiliation{Department of Mechanical Engineering, Virginia Tech, Blacksburg, Virginia 24061}

\date{\today}

\begin{abstract}
We numerically study the propagation of reacting fronts in a shallow and horizontal layer of fluid with solutal feedback and in the presence of a thermally driven flow field of counter-rotating convection rolls.  We solve the Boussinesq equations along with a reaction-convection-diffusion equation for the concentration field where the products of the nonlinear autocatalytic reaction are less dense than the reactants. For small values of the solutal Rayleigh number the characteristic fluid velocity scales linearly, and the front velocity and mixing length scale quadratically, with increasing solutal Rayleigh number. For small solutal Rayleigh numbers the front geometry is described by a curve that is nearly antisymmetric about the horizontal midplane.  For large values of the solutal Rayleigh number the characteristic fluid velocity, the front velocity, and the mixing length exhibit square-root scaling and the front shape collapses onto an asymmetric self-similar curve. In the presence of counter-rotating convection rolls, the mixing length decreases while the front velocity increases. The complexity of the front geometry increases when both the solutal and convective contributions are significant and the dynamics can exhibit chemical oscillations in time for certain parameter values. Lastly, we discuss the spatiotemporal features of the complex fronts that form over a range of solutal and thermal driving.

\end{abstract}

\maketitle

\section{Introduction}

Reacting fronts that propagate through a moving fluid are important parts of many systems in science and engineering that are of intense current interest~\cite{vansarloos:2003,xin:2000,tiani:2018}.  This includes geophysical problems such as the lock-exchange instability~\cite{shin:2004,bou-malham:2010} of oceanic and atmospheric flows, the buoyancy and surface tension driven flows of chemical fronts~\cite{tiani:2018,dhernoncourt:2007,rogers:2012,doan:2018,budroni:2019}, the propagation of polymerization fronts~\cite{belk:2003}, the rich spatiotemporal dynamics of forest fires~\cite{hargrove:2000,pastor:2003}, and the improved properties of combustion of pre-mixed gases in a turbulent fluid flow~\cite{williams:1985,sreenivasan:1989,sabelnikov:2015}.

In many situations of interest, the propagating front and the fluid dynamics are coupled resulting in a rich and complex dynamics. For example, the reactants and products may have different densities and the reaction may generate or absorb heat. This solutal and thermal feedback between the front and the fluid can fundamentally affect the dynamics. Furthermore, when the front propagates through an externally generated fluid velocity field, such as a turbulent flow, the interactions between reaction, convection, and diffusion contributions can become very complex.

Much of the initial interest in this problem was generated by pioneering experiments of autocatalytic reaction fronts traveling through capillary tubes at different orientations with respect to the direction of gravity~\cite{pojman:1990,pojman:1991,pojman:1991:p3,masere:1994}.  Of particular interest was the convective flows that were driven by the reaction. This led to further experimental studies over a range of conditions including channels~\cite{rogers:2005,bou-malham:2010,popity-toth:2012,schuszter:2015}, Petri dishes~\cite{miike:2015} and Hele-Shaw cells~\cite{rongy:2009:chaos,schuszter:2009,jarrige:2010,popity-toth:2011}.

There have been several numerical investigations of propagating fronts with feedback, through an initially quiescent fluid, that are directly relevant to our study. An early investigation by Vasquez \emph{et al.}~\cite{vasquez:1994} used a two-dimensional truncated Galerkin approach valid for sharp fronts near the threshold of solutal convection for the conditions of capillary tube experiments.  This approach was used to explore the speed and shape of the front and to quantify the enhanced front velocity in the presence of any convective motion~\cite{vasquez:1994}.

Rongy~\emph{et al.} have numerically explored horizontally traveling fronts using a two-dimensional Stokes flow approximation for a wide range of conditions including solutal feedback only~\cite{rongy:2007} and for layers with solutal and thermal feedback~\cite{rongy:2009:chaos,rongy:2009:jcp}. For fronts with solutal feedback only, it was found that a measure of the mixing length and the front velocity scaled with a square-root dependence on the solutal Rayleigh number, and that the profiles of the concentration and fluid velocity exhibit self-similar features, for large values of the solutal Rayleigh number~\cite{rongy:2007}. Jarrige \emph{et al.}~\cite{jarrige:2010} used a two-dimensional lattice Bathnagar-Gross-Krook (BGK) approach to integrate gap-averaged equations in an effort to account for the no-slip sidewalls used in front propagation experiments conducted in Hele-Shaw cells.

Considerable theoretical insight has been gained using a thin front, or eikonal, description of the front that is valid when the front length scale is much smaller than the length scale of the fluid motion~\cite{edwards:1991,masere:1994,jarrige:2010,bou-malham:2010}. For the horizontal layers that we are interested in studying, this corresponds to the case where the depth of the fluid layer is much larger than the front thickness. In this case, it is possible to directly quantify the connection between the front shape, fluid velocity, and front velocity through an eikonal relation. Bou-Malham \emph{et al.}~\cite{bou-malham:2010} provide a theoretical description using the eikonal description of thin fronts with solutal feedback which yields the square root dependence of the mixing length and the front velocity with the solutal Rayleigh number.

Significant attention has been paid to the study of propagating fronts through externally generated flow fields in the absence of solutal or thermal feedback~(\emph{c.f.}~\cite{audoly:2000,abel:2001,abel:2002,doan:2018,mukherjee:2019}). In this case, an aspect of interest is the enhancement of the front velocity in the presence of imposed fluid motion. However, much less is understood for fronts with feedback traveling through convective flow fields.

In this article, we focus upon a reacting front whose products are less dense than the reactants where the front propagates horizontally with respect to gravity through a shallow layer of fluid as shown in Fig.~\ref{fig:geometry}.   We also assume that the reaction is isothermal and therefore the propagating front does not generate or remove heat. The products, being less dense than the reactants, generate fluid motion due to buoyancy.  This coupling between the concentration and the fluid flow we will refer to as solutal coupling or feedback. We emphasize that the solutal coupling is two-way in the sense that concentration changes affect the flow field which can then affect the concentration field.

The paper is organized as follows. We first explore propagating fronts with solutal feedback in the absence of thermal convection. In this case, all of the fluid motion is a result of the solutal coupling caused by the density changes due to the chemical reaction. We use this to build an understanding of the solutally driven convection roll that is formed and propagates with the front. We are particularly interested in its features for small solutal driving where we use a perturbation approach, and for large solutal driving where we examine the presence of scaling ideas. This provides insights that we then use to study fronts with solutal feedback that propagate through a field of convection rolls generated by Rayleigh-B\'enard convection. We explore the complex interplay between the fluid dynamics of the convection rolls and the fluid dynamics driven by the solutal feedback of the propagating front. We quantify the flow structures that emerge which include oscillatory dynamics. Lastly, we present some concluding remarks.

\section{Approach}

\begin{figure}[tbh]
   \begin{center}
   \includegraphics[width=3.25in]{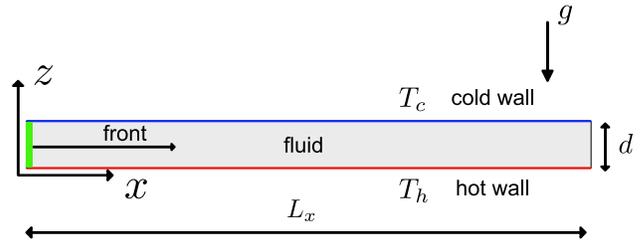} 
   \caption{(color online) The two-dimensional geometry used to study propagating fronts. The fluid layer has a depth $d$ and length $L_x$ where the bottom wall is hot (red) at temperature $T_h$ and the top wall is cold (blue) at temperature $T_c$. The coordinate directions $(x,z)$ are shown where $z$ opposes gravity $g$. The aspect ratio is $\Gamma\!=\!L_x/d$ and the front is initiated at the left wall (green) and propagates to the right in the $x$ direction. The domain illustrated here is not to  scale, in the numerical simulations $\Gamma=30$ unless stated otherwise.}
    \label{fig:geometry}
    \end{center}
\end{figure}

The schematic shown in Fig.~\ref{fig:geometry} illustrates the geometric details of the two-dimensional fluid layer that we explore. The shallow fluid layer has a depth $d$ and a length $L_x$ where the aspect ratio of the domain is $\Gamma = L_x/d \!\gg \!1$. The bottom surface is hot and is at temperature $T_h$ and the top surface is cold and is at temperature $T_c$ where $\Delta T = T_h - T_c$ is a constant. The $z$ direction is opposing to gravity $g$ and the front propagates in the $x$ direction. In our study, the front is always initiated at the left wall where $x\!=\!0$ and propagates to the right. A front at initiation is shown by the vertical green stripe.
%%%%%%%%%%%%%%%%%%%%%%%%%%%%%%%%%%%%%%%%%%%%
%%%%%%%%%%%%%%%%%%%%%%%%%%%%%%%%%%%%%%%%%%%%

The governing equations are determined by applying the conservation of momentum, energy, mass, and chemical species to yield
\begin{eqnarray}
\text{Pr}^{-1} \! \left( \frac{\partial \vec u}{\partial t} \!+\! \vec u \cdot \vec \nabla \vec u \right )\! &=&\! - \vec \nabla p \!+\! \nabla^2 \vec u \!+\! \text{Ra}_T T \hat z \!+\! \text{Ra}_s c \hat z, \label{eq:momentum} \\
\frac{\partial T}{\partial t} + \vec{u} \cdot \vec \nabla T &=& \nabla^2 T, \label{eq:energy} \\
\vec{\nabla} \cdot \vec{u} &=& 0, \label{eq:mass}
\end{eqnarray}
and 
\begin{equation}
\frac{\partial c}{\partial t} + \vec u \cdot \vec \nabla c = \text{Le}\nabla^2 c + \xi f(c).
\label{eq:rad}
\end{equation}
In these equations, $\vec{u}\!=\!(u,w)$ is the two-dimensional fluid velocity vector where $u(x,z,t)$ and $w(x,z,t)$ are the $x$ and $z$ components of the fluid velocity, respectively, and $t$ is time. The fluid pressure is $p(x,z,t)$, the fluid temperature is $T(x,z,t)$, and the concentration of the products is $c(x,z,t)$. These equations have been nondimensionalized using the depth of the fluid layer $d$ as the length scale, $\Delta T$ as the temperature scale, the thermal diffusion time $d^2/\alpha$ as the time scale where $\alpha$ is the thermal diffusivity of the fluid, $\mu \alpha / d^2$ as the pressure scale where $\mu$ is the dynamic viscosity, and the initial concentration of reactants $a_0$ as the concentration scale. Lastly, $\hat{z}$ is a unit vector in the $z$-direction.

Several nondimensional parameters appear in Eqs.~(\ref{eq:momentum})-(\ref{eq:rad}). The Prandtl number $\text{Pr} = \nu/\alpha$ is the ratio of diffusivities of momentum and heat. Since variations in temperature and variations in the concentration due to the reaction can alter the density of the fluid we have two Rayleigh numbers $\text{Ra}_T$ and $\text{Ra}_s$. The thermal Rayleigh number $\text{Ra}_T\!=\!\beta_T g \Delta T d^3/(\alpha \nu)$ captures the variation in density due to temperature changes where $\beta_T\!=\!-\frac{\partial \rho}{\partial T}$ is the coefficient of thermal expansion. The critical value of the thermal Rayleigh number is $\text{Ra}_c\!\simeq\!1707.6$ for an infinite layer of fluid with no-slip boundaries at the walls~\cite{cross:1993}. For $\text{Ra}_T\!\ge\!\text{Ra}_c$ there will be fluid motion over the entire layer of fluid due to the thermal convective instability. We will use a supercritical thermal Rayleigh number $\text{Ra}_T > \text{Ra}_c$ to generate a convective flow field of counter-rotating rolls upon which the reacting front will propagate through.

The solutal Rayleigh number $\text{Ra}_{s}\!=\!\beta_{s} g a_0 d^3/(\alpha \nu)$ describes the variation in density with changes in concentration where $\beta_s\!=\!-\frac{\partial \rho}{\partial c}$ is the coefficient of expansion due to changes in chemical composition. It is important to highlight that there will be convective motion for \emph{any} nonzero value of $\text{Ra}_s$. This is because a vertical front, propagating horizontally and perpendicular to the gravitational field is always unstable to a density difference between the products and the reactants. The more dense species will always go under the less dense species as the front propagates in an instability that is often referred to as a lock-exchange instability which is an important component of many geophysical flows~\cite{bou-malham:2010,tiani:2018}.

We numerically explore the case where $\text{Ra}_s \!>\! 0$ which corresponds to products that are less dense than the reactants.  The case where $\text{Ra}_s \!<\! 0$ can be related to our results for $\text{Ra}_s \!>\! 0$ by the reflection symmetry about the $z\!=\!1/2$ midplane~\cite{rongy:2007}. We note that this reflection symmetry is also present for the fronts we study through counter-rotating convection rolls.

For the reaction term $f(c)$ we use the Fisher-Kolmogorov-Petrovsky-Piskunov (FKPP) nonlinearity~\cite{fisher:1937,kolmogorov:1937} which is used to model a broad range of reactions and phenomena~\cite{vansarloos:2003,cencini:2003}. This autocatalytic chemical reaction is described using the quadratic expression $f(c)\!=\!c (1\!-\!c)$. In this case, $\xi\!=\!\tau_\alpha/\tau_r$ is the ratio of the thermal diffusion time $\tau_\alpha\!=\!d^2/\alpha$ to the reaction time scale $\tau_r\!=\!(k_r a_0^2)^{-1}$ where $k_r$ is the rate constant of the autocatalytic reaction. Lastly, the Lewis number $\text{Le} = D/\alpha$ is the ratio of the mass and thermal diffusivities. 

Equations~(\ref{eq:momentum})-(\ref{eq:mass}) have used the Boussinesq approximation which assumes a linear variation of the density with changes in temperature and in concentration.  As a result, and following the approach described in~\cite{rongy:2009:jcp}, the concentration and temperature dependent density $\rho(c,T)$ can be expressed as
\begin{equation}
\rho(c,T)\!=\!-\text{Ra}_s c - \text{Ra}_T T.
\end{equation}
The nondimensional density $\rho$ is defined as $\rho = (\rho^* - \rho_0)/\rho_c$ where $\rho^*$ is the dimensional density, $\rho_0$ is the reference density, and $\rho_c = \mu \alpha/(d^3 g)$ is the characteristic scale used for the density. The reference density $\rho_0$ is the dimensional density in the absence of thermal or concentration gradients and the characteristic density $\rho_c$ is the density scale given by the pressure scale divided by the product of the length scale with gravity.  Using this description, pure reactants ($c\!=\!0$) that are cold ($T\!=\!0$) have a nondimensional density of $\rho\!=\!0$ and the density becomes negative $\rho\!<\!0$ in the presence of a temperature increase or due to changes in composition caused by the reaction.

At all material boundaries we use the no-slip boundary condition $\vec{u}\!=\!0$ for the fluid and a no-flux boundary condition $\vec{\nabla} c \cdot \hat{n}\!=\!0$ for the concentration field where $\hat{n}$ is an outward pointing unit normal. The bottom plate at $z\!=\!0$ is hot and is held at constant temperature $T\!=\!1$ and the top plate at $z\!=\!1$ is cold and is held at a constant temperature of $T\!=\!0$. The lateral sidewalls at $x=0$ and $x=\Gamma$ are perfect thermal conductors. The initial condition for the concentration profile $c(x,z,t=0)$ is chosen to be sufficiently steep to generate a pulled front. Specifically, we use  $c(x,z,t\!=\!0)\!=\!e^{-(\xi/\text{Le})^{1/2} x}$, the necessary steepness conditions  are described in detail in~\cite{vansarloos:2003}.

For simulations in the absence of a background convection flow field, the initial conditions are no fluid velocity. For our investigation of fronts propagating through a convective flow field, we first perform a long-time numerical simulation for a supercritical Rayleigh number in order to generate a field of counter-rotating convection rolls.

In general, and unless stated otherwise, we have used the following parameters in our numerical simulations. The long and shallow two-dimensional domain has an aspect ratio of $\Gamma \!=\! 30$ and the fluid has a Prandtl number of $\text{Pr}\!=\!1$ and a Lewis number of $\text{Le}\!=\!0.01$. When we include thermal convection we have used a thermal Rayleigh number of $\text{Ra}_T \!=\! 3000$ to generate a time independent chain of counter-rotating convection rolls. For the nonlinear autocatalytic reaction we have used a nondimensional reaction rate of $\xi \!=\! 9$. We have conducted simulations over the range of solutal Rayleigh numbers $0 \! \le \! \text{Ra}_s \! \le \! 8000$.

Equations~(\ref{eq:momentum})-(\ref{eq:rad}) are integrated forward in time using the high-order, parallel, and open-source spectral element solver nek5000~\cite{nek5000}. The spectral element approach is exponentially convergent in space and third-order accurate in time. High spatial resolution was required in order to capture the intricate features of the propagating fronts. We used 480 equally-sized square spectral-elements with 20$^{\text{th}}$ order interpolation polynomials. We performed spatial and temporal convergence tests to ensure the accuracy of our results. This approach has been used to explore a wide variety of problems in fluid dynamics including Rayleigh-B\'enard convection~\cite{paul:2003}, propagating fronts in chaotic flow fields without feedback~\cite{mehrvarzi:2014,mukherjee:2019}, and turbulent convection~\cite{scheel:2013} to name only a few.

\section{Results and Discussion}

\subsection{A Propagating Front with Solutal Feedback}

\begin{figure}[tbh]
   \begin{center}
   \includegraphics[width=3.4in]{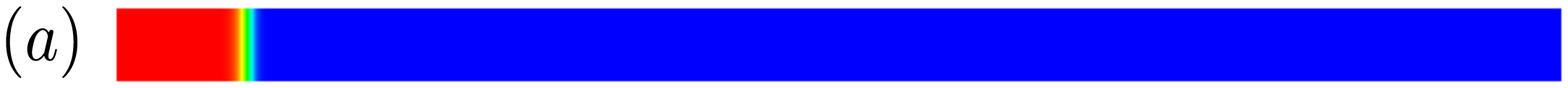} \\ 
   \includegraphics[width=3.4in]{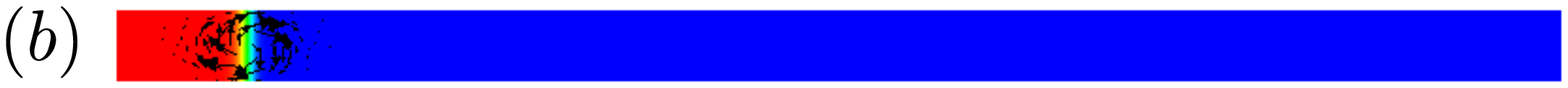} \\ 
   \includegraphics[width=3.4in]{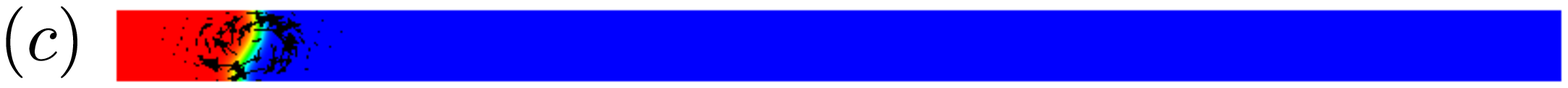} \\ 
   \includegraphics[width=3.4in]{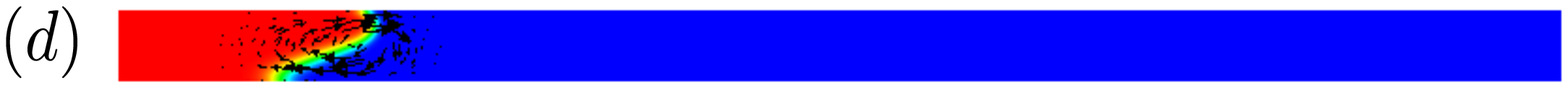} \\ 
   \includegraphics[width=3.4in]{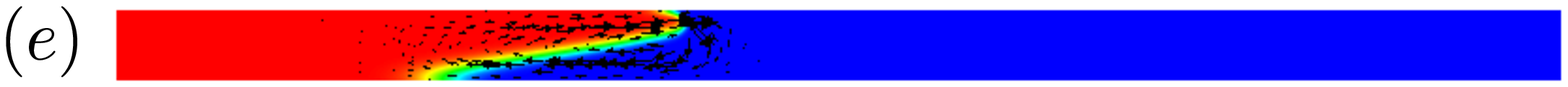} \\ 
   \includegraphics[width=3.4in]{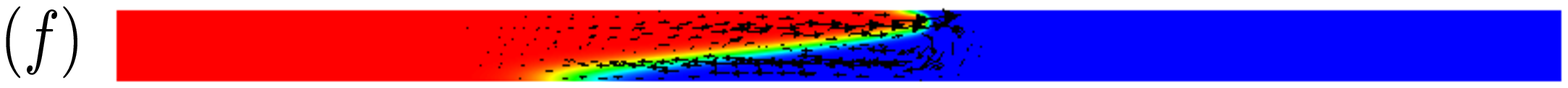} \\ 
   \includegraphics[width=3.4in]{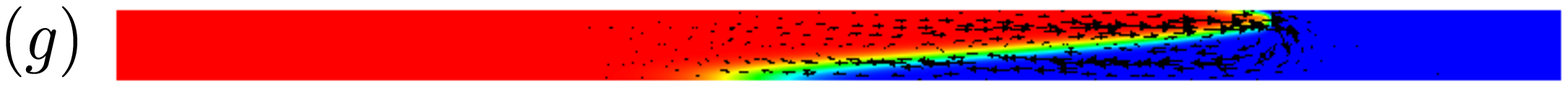} \\ 
   \includegraphics[width=3.4in]{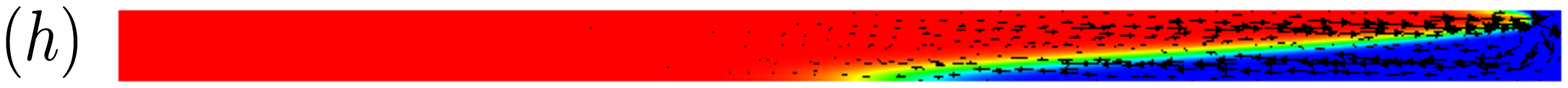}  \\ 
   \vspace{0.2cm}
   \includegraphics[width=0.75in]{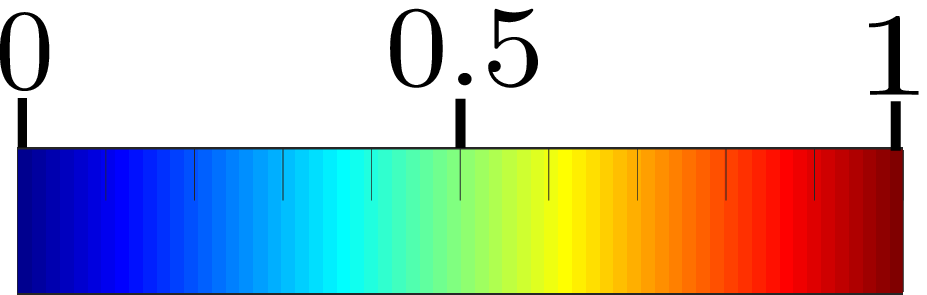} 
   \caption{(color online) Fronts propagating through an initially quiescent fluid (a)~without solutal feedback $\text{Ra}_s\!=\!0$ and~(b)-(h) with solutal feedback $\text{Ra}_s \! > \! 0$. Color contours are of the concentration c where red is pure products $(c\!=\!1)$, blue is pure reactants $(c\!=\!0)$, and green and yellow regions indicate the reaction zone or front. The front is traveling from left to right. The arrows are the fluid velocity vectors generated by the front through solutal feedback. Only a portion of the layer is shown where the left boundary is at $x\!=\!4$ and the right boundary is at $x\!=\!21.5$. For all panels $t\!=\!5$ and $\text{Ra}_T\!=\!0$. (a)-(h):~$\text{Ra}_s\!=\!\{0,0.1,10,100,500,1000,2000,3000\}$.}
    \label{fig:fronts-solutal-feedack-only}
    \end{center}
\end{figure}

We first explore propagating fronts with solutal feedback through an initially quiescent fluid layer. Figure~\ref{fig:fronts-solutal-feedack-only} illustrates several fronts over the range of solutal Rayleigh numbers $0 \!\le\! \text{Ra}_s \!\le \!3000$ where $\text{Ra}_T\!=\!0$. The images of the front and fluid motion are representative of the asymptotic state where the front has a fixed shape and propagates toward the right at a constant velocity.  The color contours are of the concentration $c(x,z,t)$ where red is pure products ($c\!=\!1$), blue is pure products ($c\!=\!0$), and the yellow and green region is the front or reaction zone. In all cases, the front is initiated at the far left and propagates to the right.  Each panel shows $4 \! \le \!x \! \le \! 21.5$, the actual domain used in the simulations is larger.  The arrows are vectors of the fluid velocity that is generated by the solutal feedback.

Figure~\ref{fig:fronts-solutal-feedack-only}(a) shows a front without solutal feedback $\text{Ra}_s\!=\!0$. In this case, the front remains vertical, there is no generation of fluid motion, and the front velocity is given by $v_0 \!=\! 2\sqrt{\text{Le} \xi} \!=\! 0.6$~\cite{vansarloos:2003}. Panels (b)-(h) are for increasing values of $\text{Ra}_s$. For $\text{Ra}_s \!>\! 0$, a self-organized solutally induced convection roll is formed with a clockwise rotation that propagates with the front. All images are at time $t\!=\!5$ where the front was initiated at $t\!=\!0$ and the relative location of the fronts indicate that the front velocity increases with increasing $\text{Ra}_s$. As $\text{Ra}_s$ increases, the front tilts to the right, is stretched over a larger distance, and develops positive and negative curvature.
%%%%%%%%%%%%%%%%%%%%%%%%%%%%%%%%%%%%%%%%%%%%
%%%%%%%%%%%%%%%%%%%%%%%%%%%%%%%%%%%%%%%%%%%%

We first quantify the propagating front and the solutally induced convection roll using the mixing length $L_s$~\cite{rongy:2007}. The mixing length is a measure of the axial distance over which the reaction occurs. The mixing length is defined in terms of the vertical average of the concentration field
\begin{equation}
\langle c(x,t) \rangle = \int_0^1 c(x,z,t) dz.
\label{eq:cave}
\end{equation}
This average value of the concentration is nearly zero at the leading edge of the front (farthest to the right) and is nearly unity at the trailing edge (farthest to the left). We follow Ref.~\cite{rongy:2007} and define $L_s(t)$ as the distance between $x$-locations where $\langle c(x,t) \rangle \!=\! 0.01$ and $\langle c(x,t) \rangle \!=\! 0.99$. We will refer to the long-time asymptotic value of $L_s(t)$ as $\bar{L}_s$.

In the absence of solutal feedback, the bare front thickness $L_0$ can be estimated as $L_0 \!=\! \bar{L}_s(\text{Ra}_s\!=\!0)$. This yields $L_0\!=\!0.598$ which is also illustrated by the width of the green and yellow vertical stripe shown in Fig.~\ref{fig:fronts-solutal-feedack-only}(a).  An important parameter that is useful in the determination of the regime of the front dynamics is the ratio $\Gamma_r$ of the thickness of the fluid layer to the bare front thickness~\cite{jarrige:2010}. $\Gamma_r \rightarrow 0$ is the mixing regime and $\Gamma_r \gtrsim 1000$ is the eikonal regime where the front is sharp and thin~\cite{jarrige:2010}. Using our nondimensionalization, this can be represented as $\Gamma_r\!=\!L_0^{-1} \!\approx\! 2$ where the nondimensional layer thickness is unity. As a result, the fronts we study are neither in the mixing or strongly eikonal regimes.

\begin{figure}[tbh]
   \begin{center}
   \includegraphics[width=2.25in]{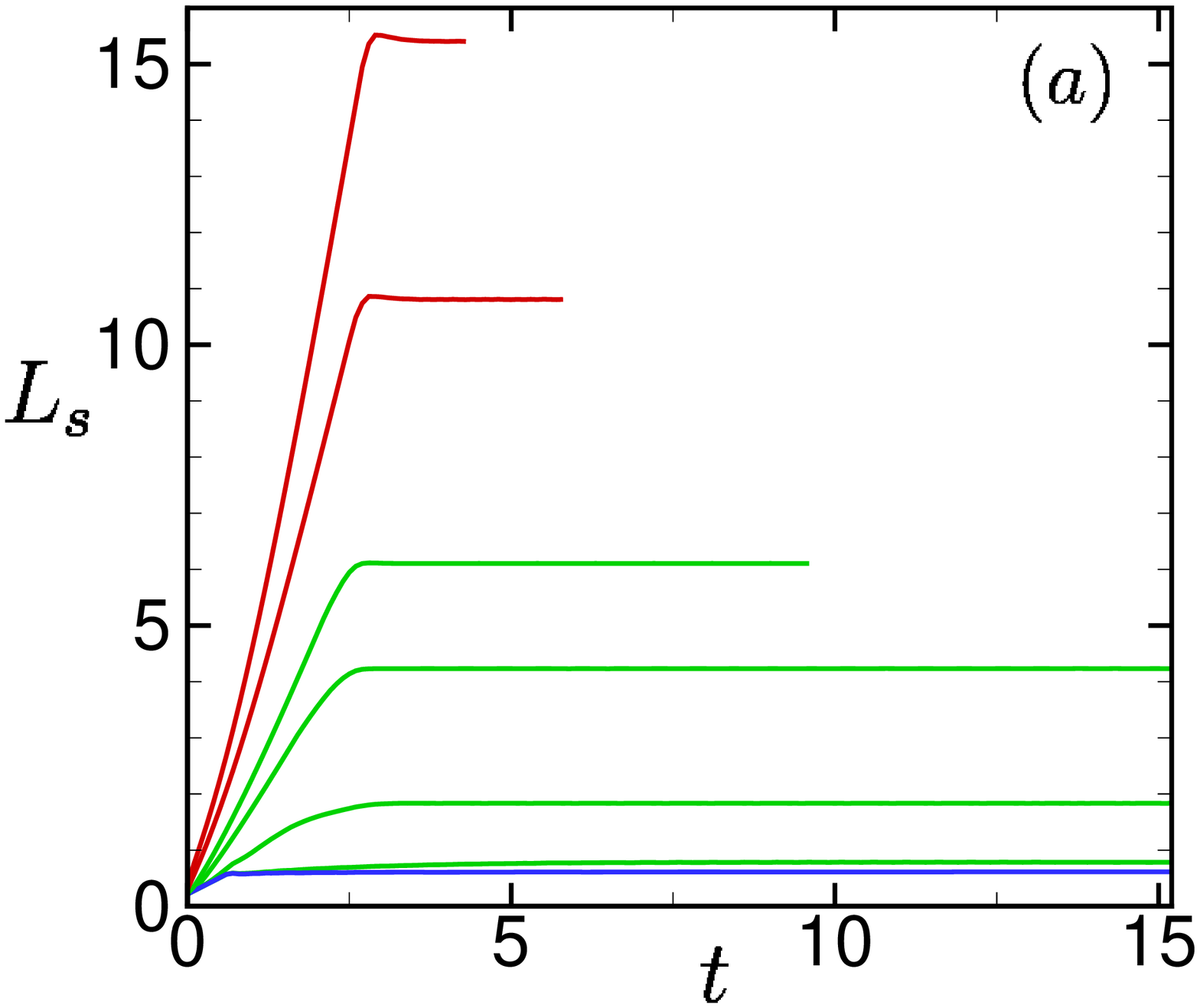} \\ 
   \includegraphics[width=1.65in]{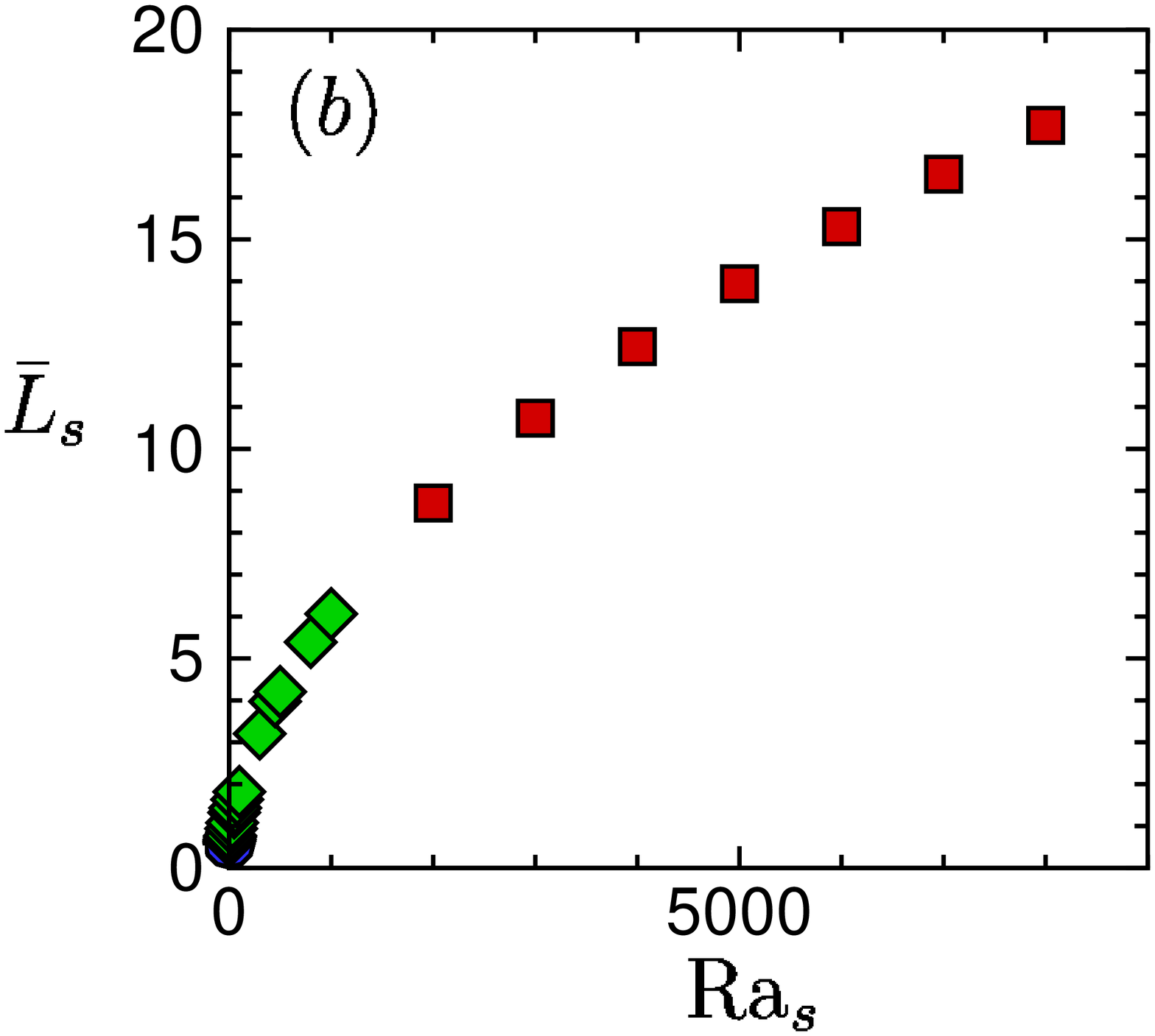} 
   \includegraphics[width=1.65in]{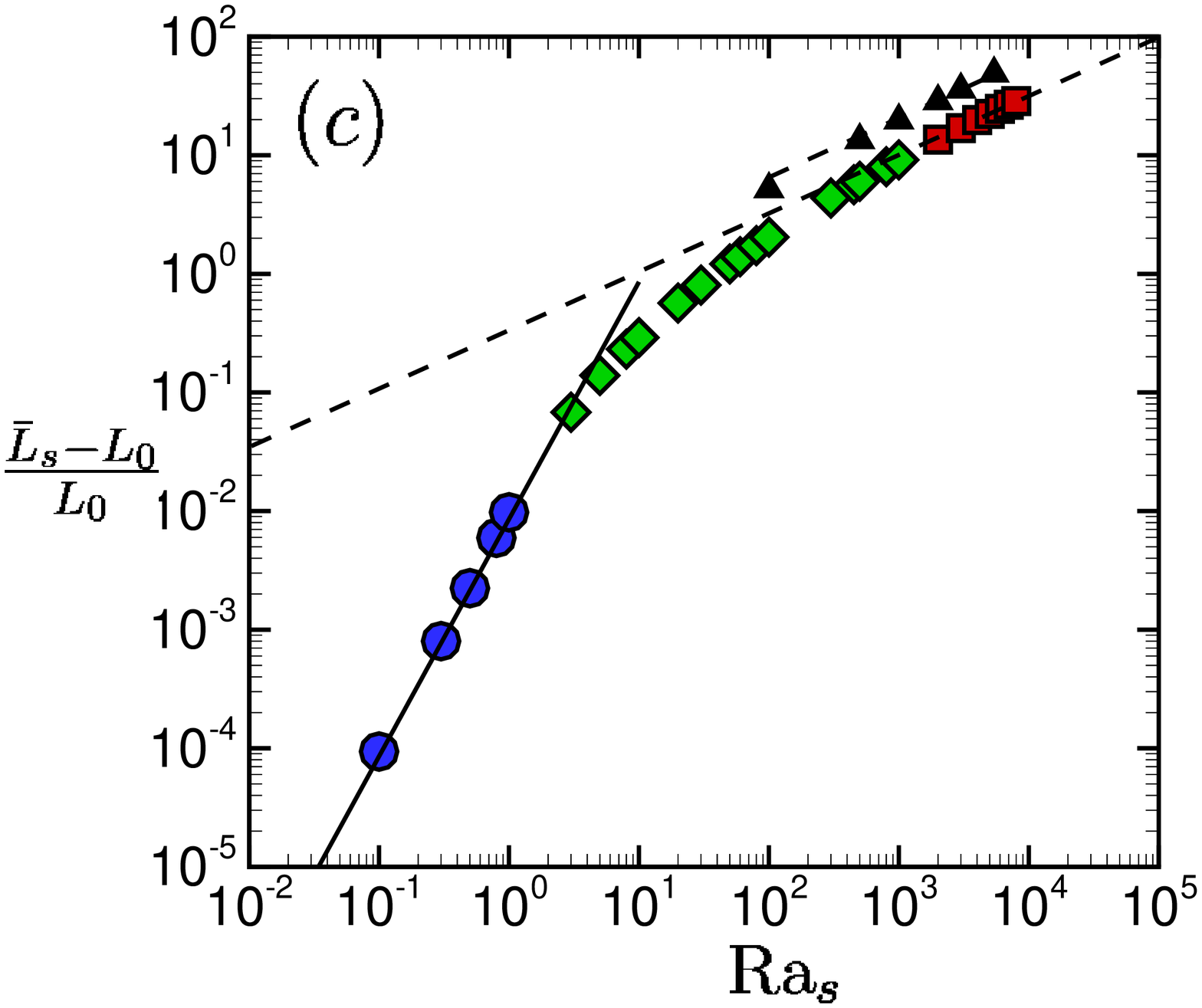} 
   \caption{(color online) The variation of the mixing length $L_s$ for $\text{Ra}_s\!>\!0$ and $\text{Ra}_T\!=\!0$. Examples of front images are shown in Fig.~\ref{fig:fronts-solutal-feedack-only}. (a)~The variation of $L_s$ with time $t$ for $\text{Ra}_s \!=\! \{1, 10, 100, 500, 1000, 3000, 6000\}$. (b)~The variation of $\bar{L}_s$ with $\text{Ra}_s$. (c)~The variation of the scaled mixing length with $\text{Ra}_s$ where $L_0\!=\!\bar{L}_s(\text{Ra}_s\!=\!0)=\!0.598$. The solid line indicates $\bar{L}_s \propto \text{Ra}_s^2$  for $\text{Ra}_s\!\le\!1$ and the dashed lines indicate $\bar{L}_s\!\propto\!\text{Ra}_s^{1/2}$ for $\text{Ra}_s\!>\!1000$.  The black triangles are results using a cubic autocatalytic reaction.}
    \label{fig:solutal-roll}
    \end{center}
\end{figure}

The time variation of $L_s(t)$ is shown in Fig.~\ref{fig:solutal-roll}(a). Each curve illustrates the mixing length as a function of time for different values of $\text{Ra}_s$. In general, $\bar{L}_s$ increases monotonically with increasing $\text{Ra}_s$. The result for $\text{Ra}_s\!=\!6000$ (the top curve in Fig.~\ref{fig:solutal-roll}(a)) yielded $\bar{L}_s \!= \!15.30$ which required a larger domain of aspect ratio $\Gamma \!=\! 60$ in order to compute the asymptotic results.

We will find it useful to discuss the results in terms of $\text{Ra}_s$ that we separate into the three ranges of low, intermediate, and large where: $0\!\le\!\text{Ra}_s \!\le\! 1$ is low, blue, and uses circles; $1\! <\! \text{Ra}_s\! \le \!1000$ is intermediate, green, and uses diamonds; and $1000 \!<\! \text{Ra}_s \!\le\! 8000$ is large, red, and uses squares. We will use this convention, color scheme, and symbol choice in all of the upcoming plots where useful.
%%%%%%%%%%%%%%%%%%%%%%%%%%%%%%%%%%%%%%%%%%%%
%%%%%%%%%%%%%%%%%%%%%%%%%%%%%%%%%%%%%%%%%%%%

Figure~\ref{fig:solutal-roll}(b)-(c) illustrates the variation of $\bar{L}_s$ with $\text{Ra}_s$. For positive values of $\text{Ra}_s$, the front tilts to the right and stretches which results in the increase in $\bar{L}_s$ as shown in Fig.~\ref{fig:fronts-solutal-feedack-only}(b)-(h). In Fig.~\ref{fig:solutal-roll}(c) we show the same results on a log-log plot where the mixing length has been normalized using $L_0$. For small values of $\text{Ra}_s$, the normalized mixing length scales quadratically as $(\bar{L}_s\!-\!L_0)/L_0 \! \propto \!  \text{Ra}_s^2$ which is indicated by the solid line. 

For large values of $\text{Ra}_s$, the results follow the square root scaling given by $(\bar{L}_s\!-\!L_0)/L_0\!=\!0.316 \text{Ra}_s^{1/2}$ which is indicated by the dashed line. For reference, we have also included results using a cubic nonlinearity for the reaction, $f(c)\!=\!c^2(1-c)$, where it is also found to exhibit the square root scaling in agreement with previous findings~\cite{rongy:2007}. The green diamonds indicate the presence of a transition region between these two scalings at small and large values of $\text{Ra}_s$.

\begin{figure}[tbh]
   \begin{center}
   \includegraphics[width=1.65in]{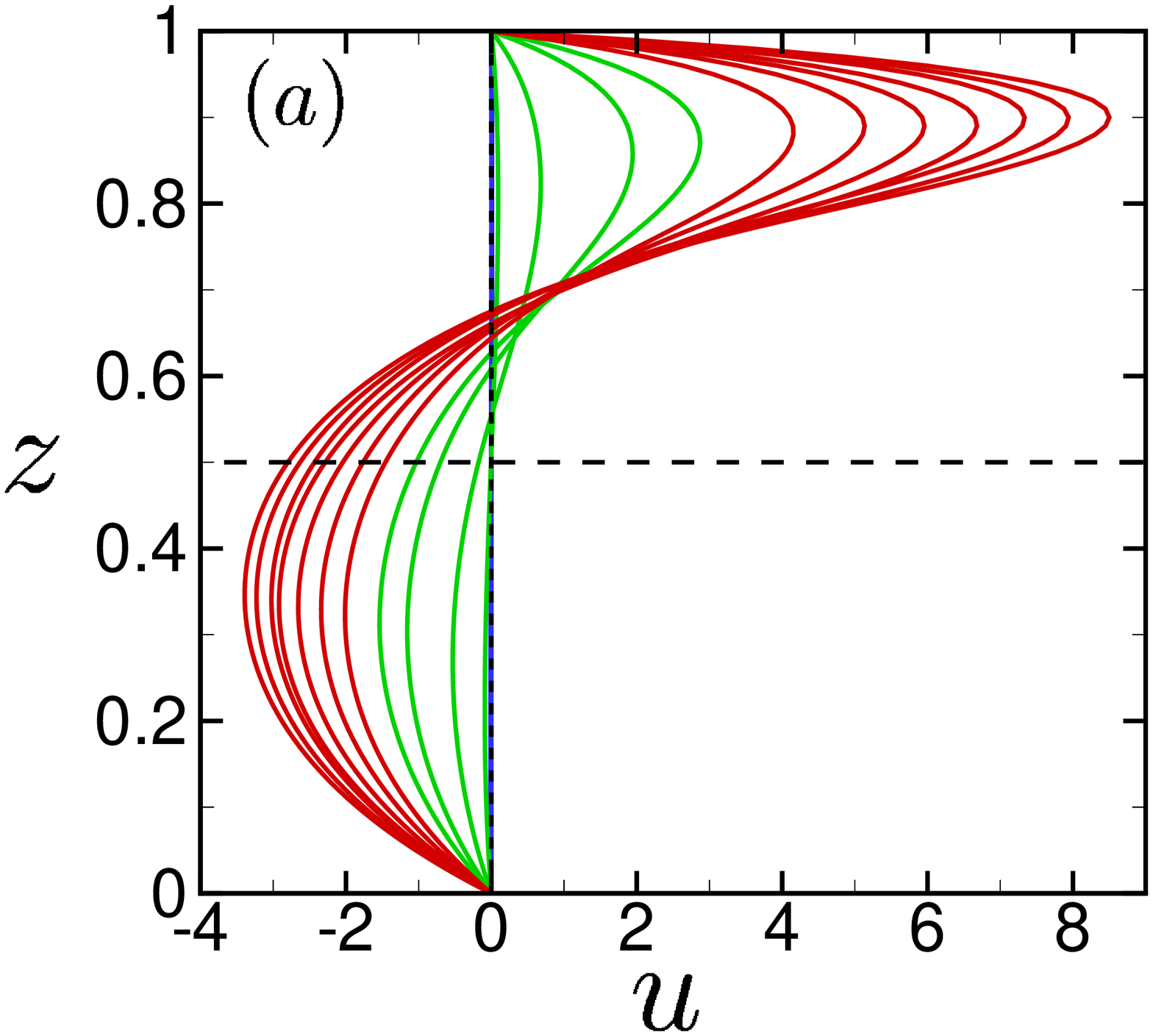}
   \includegraphics[width=1.65in]{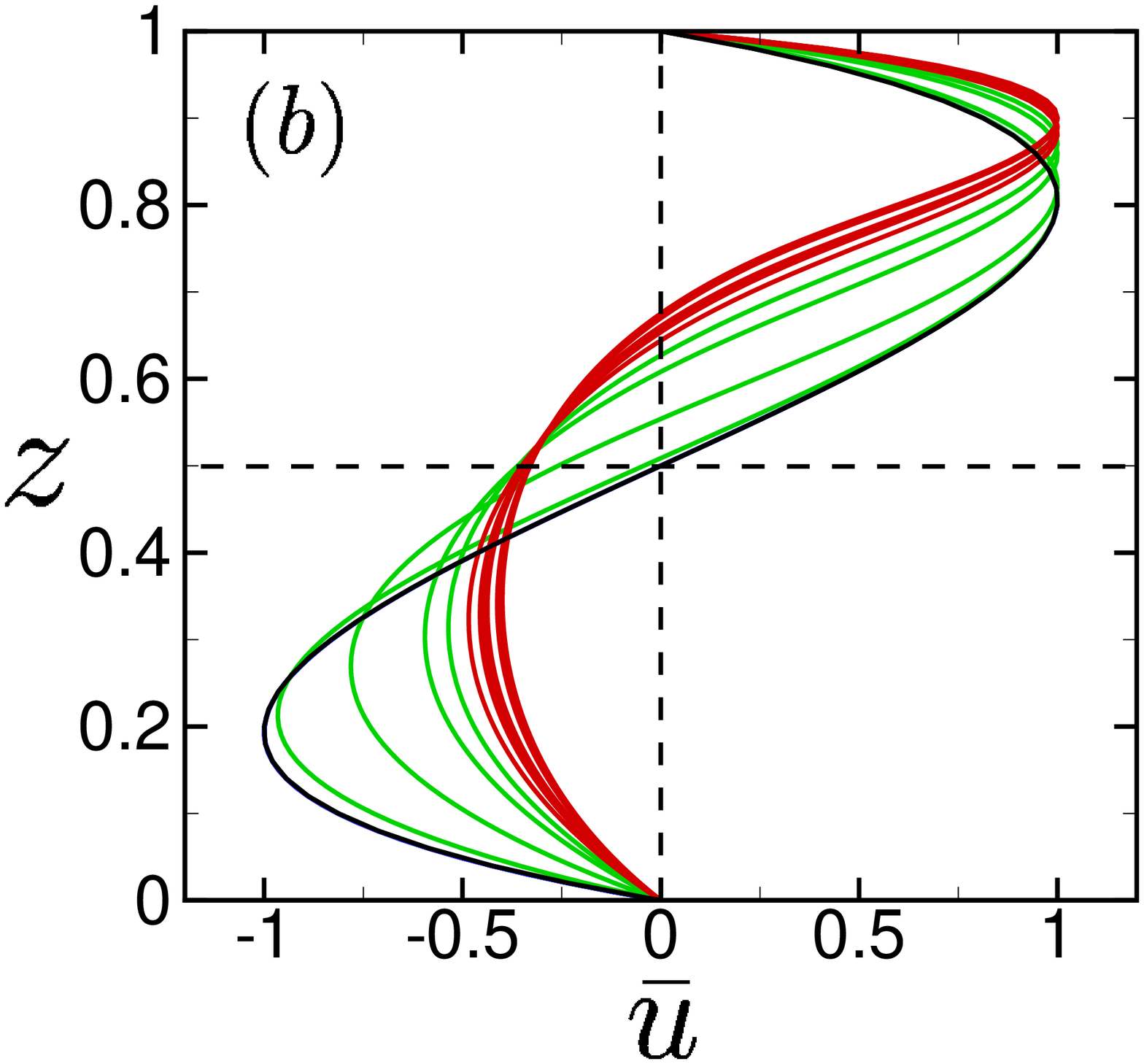} \\ 
   \includegraphics[width=1.65in]{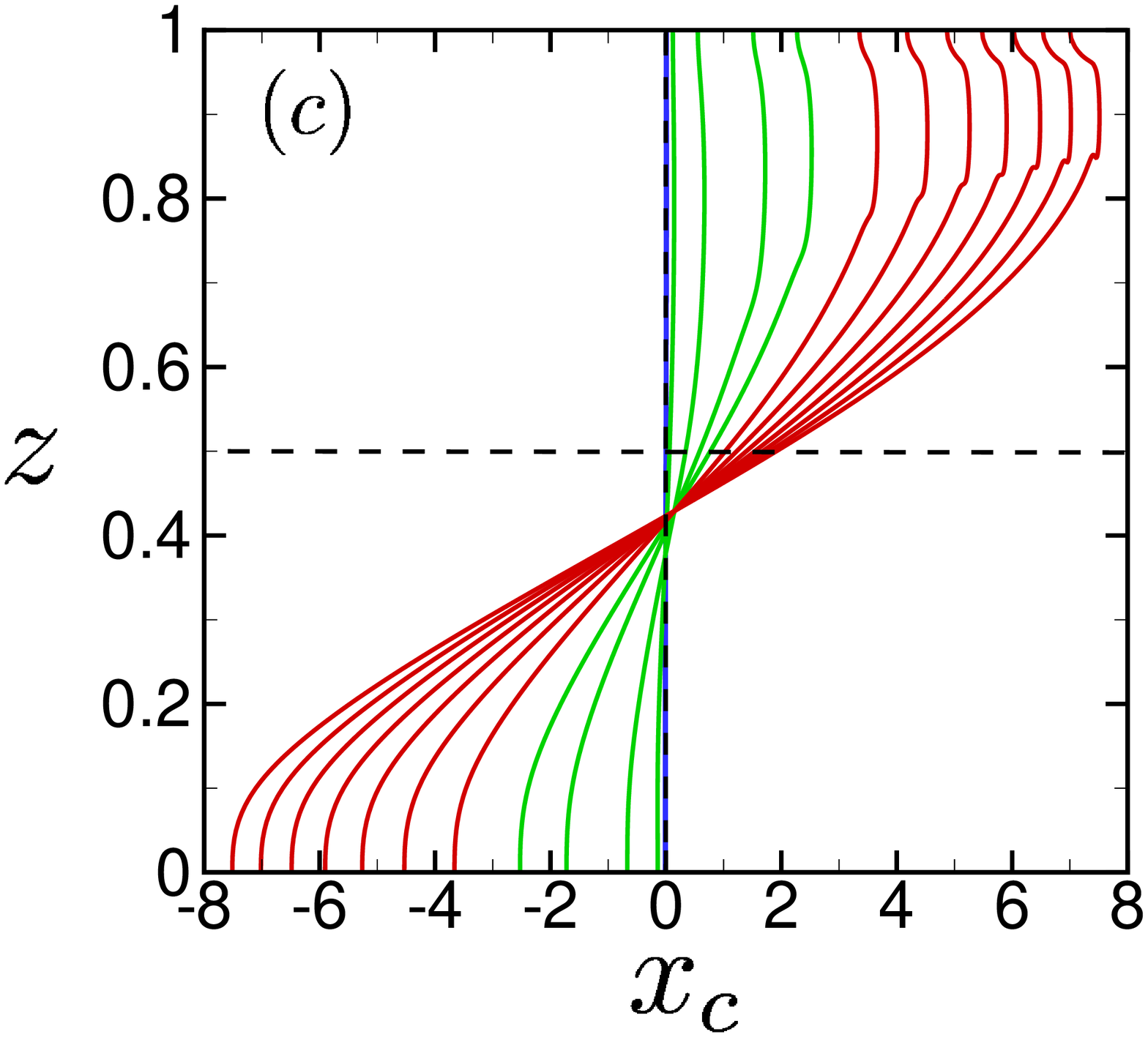} 
   \includegraphics[width=1.65in]{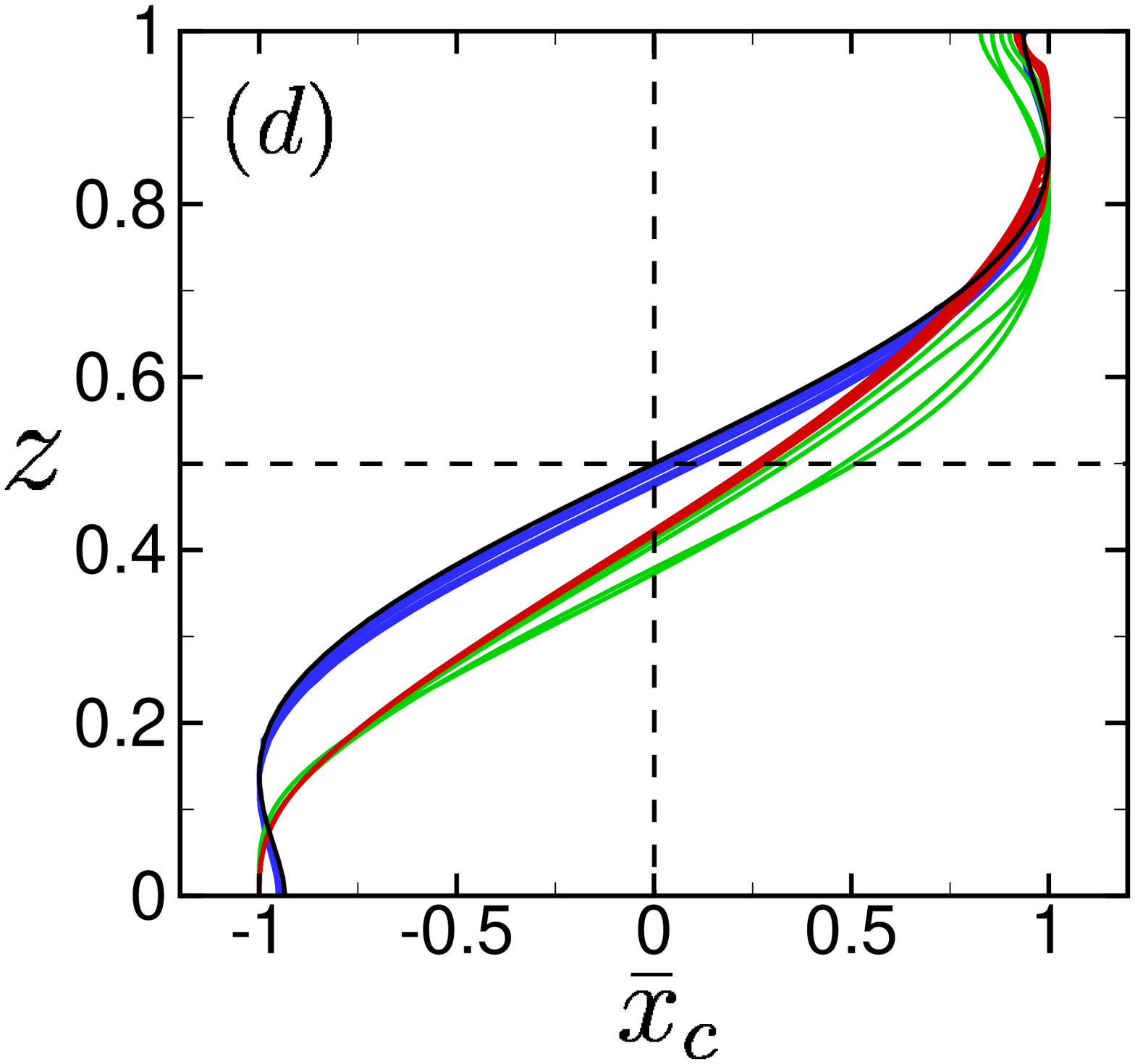} 
   \caption{(color online) Self-similar features of the front and fluid flow field in the presence of solutal feedback. All fronts have reached their asymptotic velocity and shape. The blue, green, and red curves are for small, intermediate, and large values of $\text{Ra}_s$ where $0 \! \le \! \text{Ra}_s \! \le \! 1$ (blue), $1 \! < \! \text{Ra}_s \! \le \! 1000$ (green), $1000 \! < \! \text{Ra}_s \! \le \! 8000$ (red).  Images of the fronts are in Fig.~\ref{fig:fronts-solutal-feedack-only}. (a)~The variation of the axial fluid velocity $u$ with the vertical coordinate $z$. The slice in the $z$ direction is taken at the $x$ location where $u$ is at its maximum value $u_{\text{max}}$. (b)~The same data plotted as a function of the normalized axial velocity $\bar{u}=u/u_\text{max}$. (c)~The variation of the front shape where the front is plotted as the isocontour where  $c\!=\!1/2$.  The fronts are centered using $x_c$ where $x_c=0$ is the center location of the front. (d)~The normalized front shapes using the scaled coordinate $\tilde{x}_c$. The black curves in (b)-(d) are for $\text{Ra}_s=10^{-3}$ which have been computed using a perturbation approach.}
    \label{fig:axial-flow-profiles}
    \end{center}
\end{figure}

The variation of the horizontal fluid velocity $u$ with $z$ is shown in Fig.~\ref{fig:axial-flow-profiles}(a)-(b). Each curve is $u(x,z,t)$ where the location $x$ is chosen such that the horizontal fluid velocity includes the maximum value present in the flow field at that time $t$. As a result, the position $x$ is chosen near the leading edge of the front where the fluid velocity of the solutally induced convection roll is largest.

Figure~\ref{fig:axial-flow-profiles}(a) shows profiles of $u$ for $0 \! \le \! \text{Ra}_s \! \le \! 8000$. As described by Rongy~\emph{et al.}~\cite{rongy:2007} these curves yield a self-similar description at large $\text{Ra}_s$ when the fluid velocity is scaled by its maximum value $u_\text{max}$. Our results also indicate this scaling as shown by the red curves in Fig.~\ref{fig:axial-flow-profiles}(b).

In addition, we find a self-similar structure to the flow field at small $\text{Ra}_s$ which is shown by the blue curves. The fluid velocity contours for the intermediate values of $\text{Ra}_s$ do not collapse onto a single curve and represent the transition between the low and high $\text{Ra}_s$ results. The horizontal and vertical dashed lines are included to illustrate the nearly antisymmetric shape of the low $\text{Ra}_s$ results about the midplane where $z\!=\!1/2$.  The asymmetry of the curves increase as $\text{Ra}_s$ is increased.
%%%%%%%%%%%%%%%%%%%%%%%%%%%%%%%%%%%%%%%%%%%%
%%%%%%%%%%%%%%%%%%%%%%%%%%%%%%%%%%%%%%%%%%%%

Figure~\ref{fig:axial-flow-profiles}(c)-(d) illustrate the shape of the front where the front has been identified as the isocontour of the concentration field where $c\!=\!1/2$. In this case, the fronts have also been centered using the coordinate $x_c$ where $x_c\!=\!x\!-\!(x_\text{max}\!+\!x_\text{min})/2$. $(x_\text{min},x_\text{max})$ are the minimum and maximum values of $x$ for the isocontour describing the front and, as a result, the center of each front is located at $x_c\!=\!0$.  Figure~\ref{fig:axial-flow-profiles}(d) shows the same results where we have scaled the front position such that the front location at the far right side is unity using $\tilde{x}_c \!=\! x_c/x_{c,\text{max}}$ where $x_{c,\text{max}}$ is the largest value of $x_c$ for each curve in Fig.~\ref{fig:axial-flow-profiles}(c). When plotted this way the fronts show a self-similar front shape for small (blue) and large (red) solutal Rayleigh numbers.

\begin{figure}[tbh]
   \begin{center}
   \includegraphics[width=1.65in]{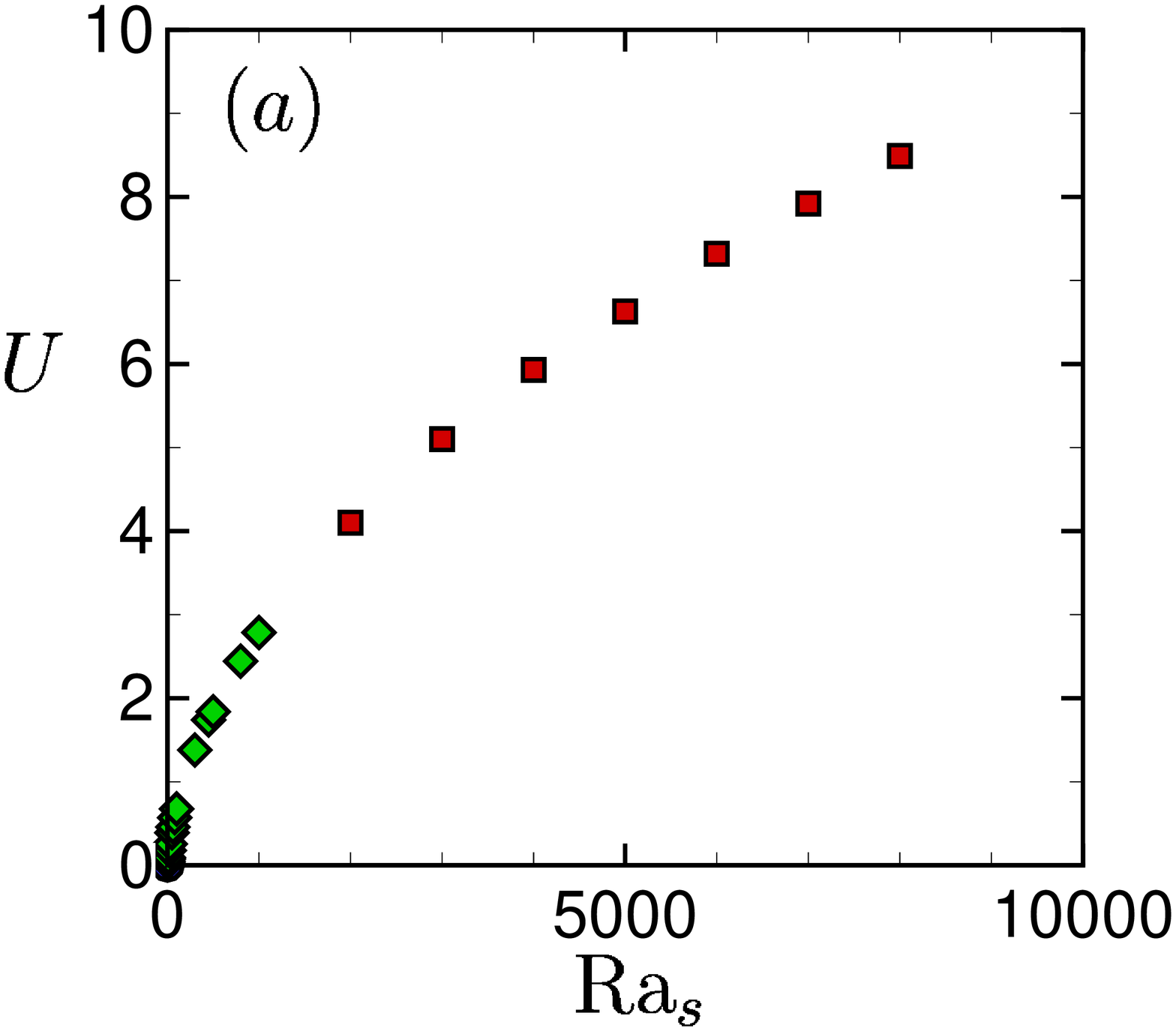}
   \includegraphics[width=1.65in]{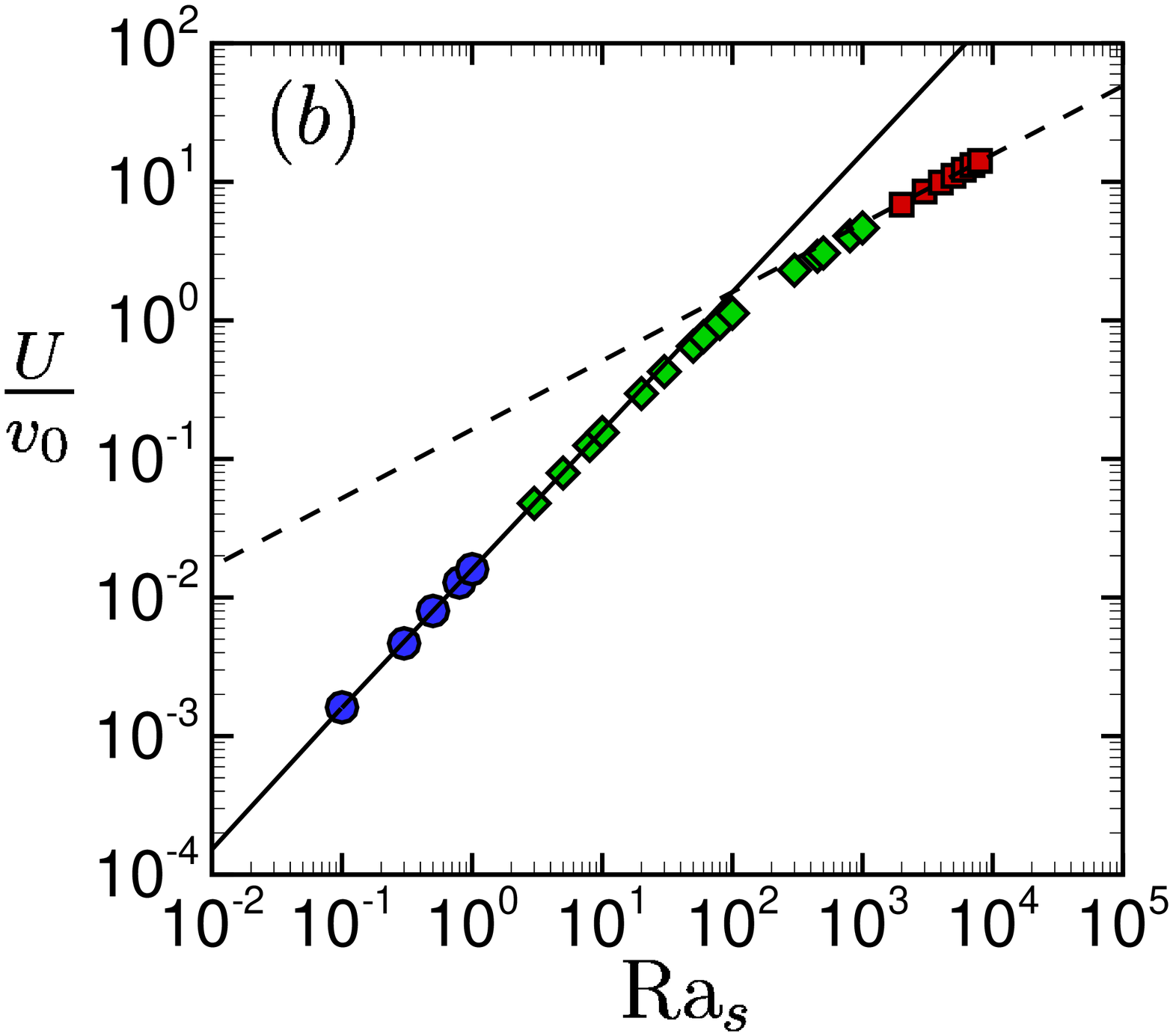} \\
   \includegraphics[width=1.65in]{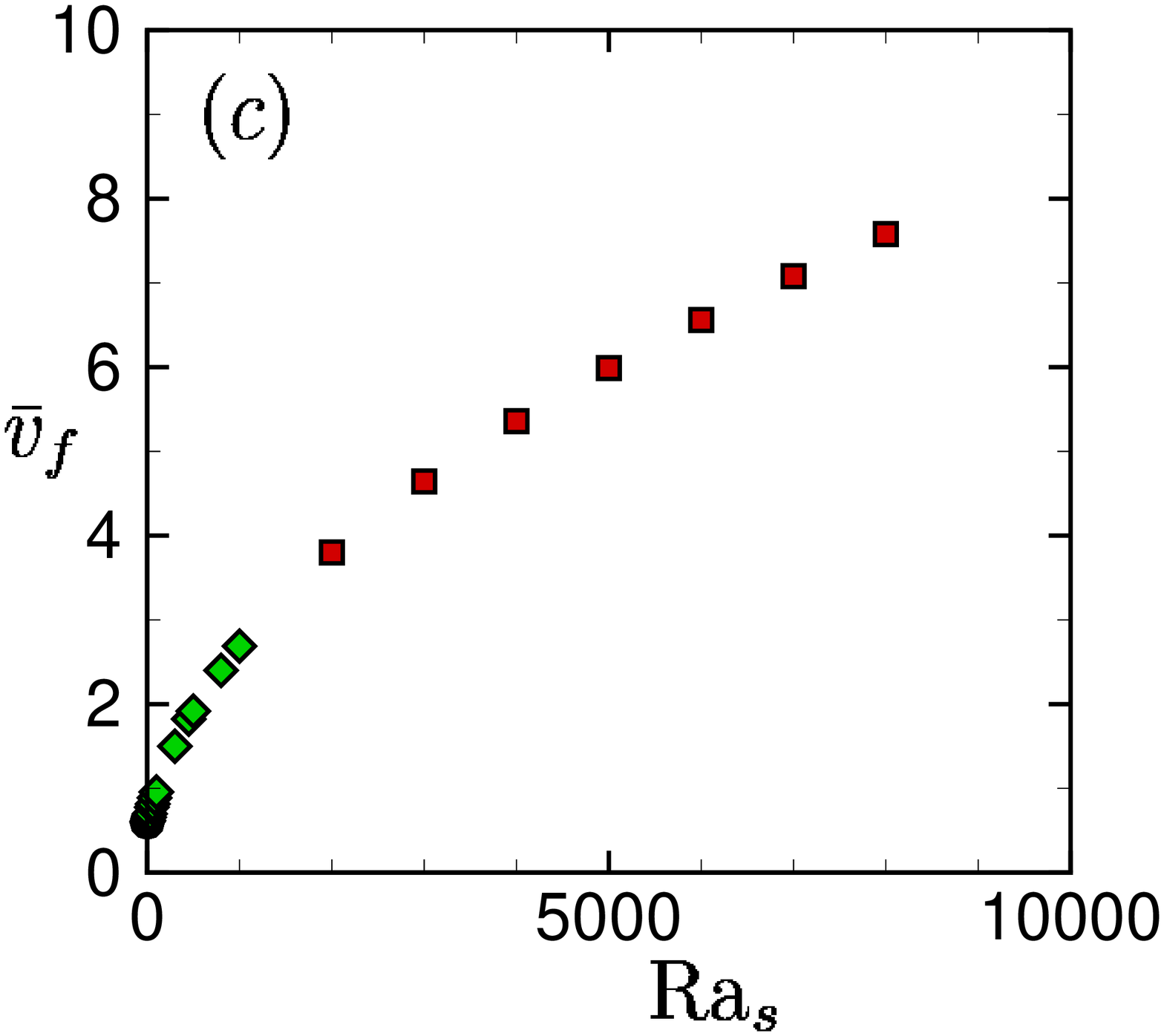} 
   \includegraphics[width=1.65in]{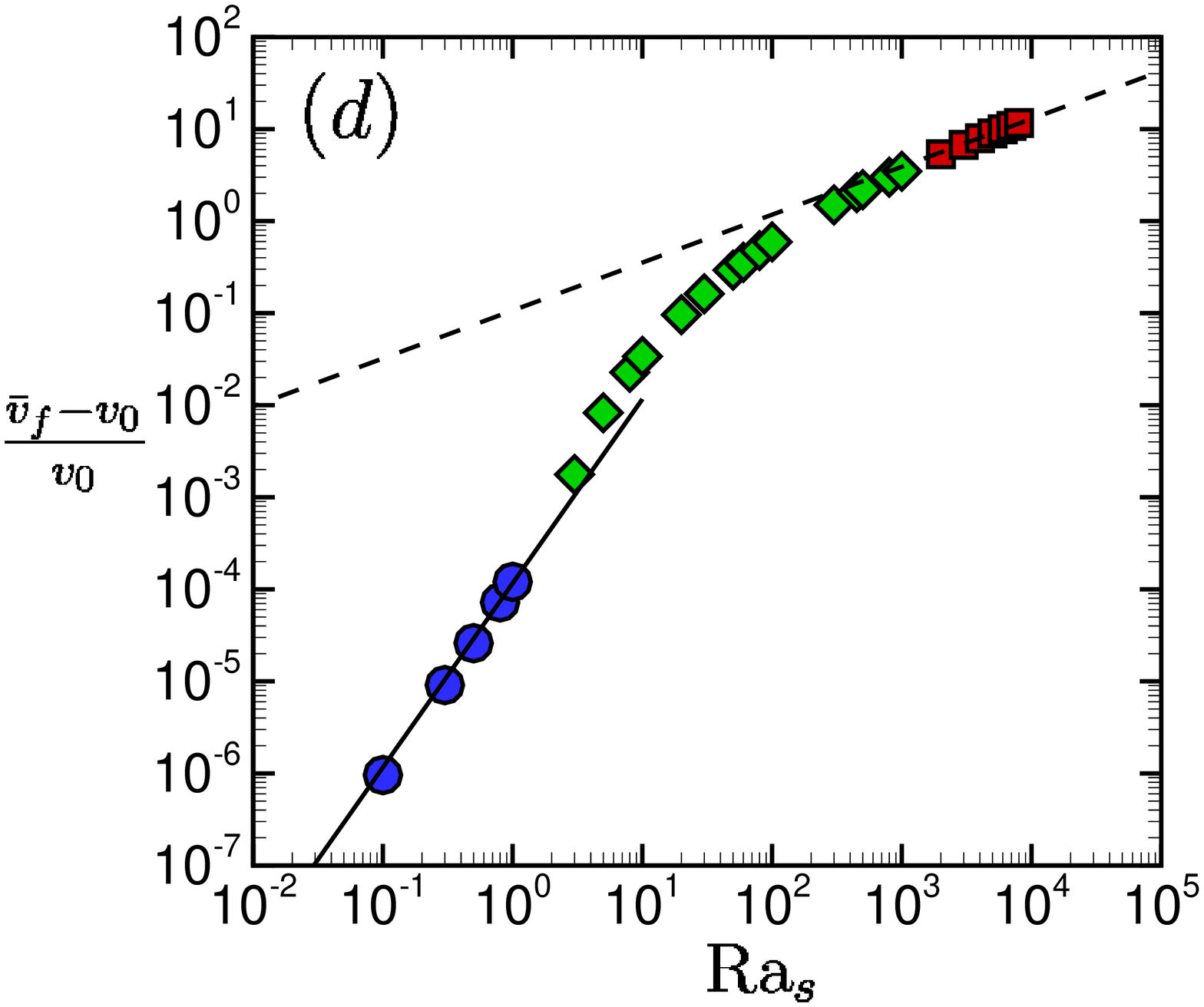} 
   \caption{The variation of the characteristic fluid velocity $U$ and the asymptotic front velocity $\bar{v}_f$ with the solutal Rayleigh number $\text{Ra}_s$ in the absence of thermal convection $\text{Ra}_T\!=\!0$. (a)~The variation of $U$ with $\text{Ra}_s$. (b)~The variation of $U/v_0$ with $\text{Ra}_s$ where $v_0$ is the bare front velocity that is found when $\text{Ra}_s\!=\!\text{Ra}_T\!=\!0$. The solid line indicates $U/v_0 \! \propto \! \text{Ra}_s$ for small $\text{Ra}_s$ and the dashed line indicates $U/v_0 \! \propto \! \text{Ra}_s^{1/2}$ for large $\text{Ra}_s$.  (c)~The variation of $\bar{v}_f$ with $\text{Ra}_s$. (d)~The variation of the scaled front velocity with $\text{Ra}_s$. The solid line indicates a $\text{Ra}_s^2$ scaling and the dashed line indicates a $\text{Ra}_s^{1/2}$ scaling. The circles (blue), diamonds (green), and squares (red) are results for small, intermediate, and large values of $\text{Ra}_s$, respectively.}
  \label{fig:fluid-and-front-velocity-no-convection}
  \end{center}
\end{figure}

Figure~\ref{fig:axial-flow-profiles}(a) illustrates that the maximum horizontal velocity of the fluid increases with increasing values of $\text{Ra}_s$ and that the location of this maximum occurs near the upper boundary. In Fig.~\ref{fig:fluid-and-front-velocity-no-convection}(a)-(b) we show how the fluid velocity scales with $\text{Ra}_s$ where $\text{Ra}_s$ varies over five orders of magnitude. To quantify the fluid motion we use the characteristic fluid velocity $U$ which is defined as the maximum value of the fluid velocity $|\vec{u}|$ over the entire domain when the front has reached its asymptotic propagating state. For fronts with $\text{Ra}_T\!=\!0$ we have $U \!\approx\! u_{\text{max}}$ where $u_\text{max}$ can be determined from Fig.~\ref{fig:axial-flow-profiles}(a). This definition of $U$ will be useful when we discuss fronts in the presence of fluid convection and the resulting fluid motion is more complex.

It is insightful to define the Reynolds number $\text{Re}$ for the flow field. Using the characteristic velocity $U$ and our nondimensionalization yields the relationship $\text{Re} = U/\text{Pr}$. In our results, $\text{Pr}\!=\!1$, and this relationship simplifies to $\text{Re}=U$. Figure~\ref{fig:fluid-and-front-velocity-no-convection}(a)-(b) indicates that for $\text{Ra}_s \lesssim 1$ the flow field is in the Stokes flow regime where $\text{Re} \ll 1$ while for the larger values of $\text{Ra}_s$ that we explore we have $\text{Re} \lesssim 10$.
%%%%%%%%%%%%%%%%%%%%%%%%%%%%%%%%%%%%%%%%%%%%
%%%%%%%%%%%%%%%%%%%%%%%%%%%%%%%%%%%%%%%%%%%%

There are several interesting trends evident in Fig.~\ref{fig:fluid-and-front-velocity-no-convection}(a)-(b). For small values of the solutal Rayleigh number $\text{Ra}_s \! \le \! 1$, shown as the blue circles, the characteristic velocity $U$ scales linearly with $\text{Ra}_s$. The linear scaling $U/v_0 \! \propto \! \text{Ra}_s$ is indicated by the solid line in Fig.~\ref{fig:fluid-and-front-velocity-no-convection}(b). The scaling then transitions to $U/v_0 \propto \text{Ra}_s^{1/2}$ for larger values where $\text{Ra}_s \ge 1000$ as shown by the red squares and the dashed line.

Figure~\ref{fig:fluid-and-front-velocity-no-convection}(c)-(d) illustrates how the asymptotic front velocity $\bar{v}_f$ varies with $\text{Ra}_s$.  In order to quantify the front velocity we use the bulk burning rate approach~\cite{constantin:2000} which can be expressed as
\begin{equation}
v_f(t) = \int_0^1 dz \int_0^\Gamma dx \frac{\partial c}{\partial t}.
\label{eq:vf}
\end{equation}
The use of the bulk burning rate for propagating fronts in chaotic flows is also described in~\cite{mukherjee:2019}.  The asymptotic value of the front velocity $\bar{v}_f$ is determined by fitting numerical results for $v_f(t)$ with $v_f(t) \!=\! \bar{v}_f \!-\! b/t$ and taking the limit of infinite time. For the fronts shown in Fig.~\ref{fig:fronts-solutal-feedack-only}, a simple front tracking approach would suffice and the result for $\bar{v}_f$ would be identical to what is found using Eq.~(\ref{eq:vf}). However, the bulk burning rate approach will be very useful when the fronts become more complicated in the presence of thermal convection where front tracking approaches become difficult to use.

Figure~\ref{fig:fluid-and-front-velocity-no-convection}(d) indicates that the scaled front velocity scales as $\text{Ra}_s^2$ for $\text{Ra}_s \!\le\! 1$ as shown by the solid line through the circles (blue).  The front velocity then transitions to a  $\text{Ra}_s^{1/2}$ scaling which is shown by the dashed line through the squares (red).

\subsection{Perturbation Analysis for $\text{Ra}_s \! \ll \! 1$}
\label{section:perturbation}

In order to gain insight into the scalings $U \propto \text{Ra}_s$, $\bar{L}_s \propto \text{Ra}_s^2$, and $\bar{v}_f \propto \text{Ra}_s^2$ at small solutal Rayleigh number we explore the problem perturbatively for $\text{Ra}_s \! \ll \! 1$.  In the following we describe the mathematical approach and the physical insights we can draw. Further details regarding the numerical approach used to solve the equations are given in the Appendix~\ref{section:numerics}.

It is convenient to first recast Eqs.~(\ref{eq:momentum})-(\ref{eq:rad}) using a stream-function vorticity formulation to remove the pressure variable and the explicit need for a separate equation for the conservation of mass of the fluid. This yields
\begin{equation}
\text{Pr}^{-1} \left( \frac{\partial \omega}{\partial t}  + \frac{\partial \psi}{\partial z} \frac{\partial \omega}{\partial x} - \frac{\partial \psi}{\partial x} \frac{\partial \omega}{\partial z} \right)  = \frac{\partial^2 \omega}{\partial x^2} + \frac{\partial^2 \omega}{\partial z^2} + \text{Ra}_{s} \frac{\partial c}{\partial x}, \label{eq:omegapsi}
\end{equation}
and
\begin{equation}
\frac{\partial c}{\partial t} +  \frac{\partial \psi}{\partial z} \frac{\partial c}{\partial x} - \frac{\partial \psi}{\partial x} \frac{\partial c}{\partial z}= \text{Le}\left(\frac{\partial^2 c}{\partial x^2} + \frac{\partial^2 c}{\partial z^2} \right) + \xi c(1-c)
\label{eq:cpsi}
\end{equation}
where $\omega(x,z,t) \!=\! (\vec{\nabla} \times \vec{u}) \cdot \hat{y}$ is the $y$-component of the fluid vorticity vector and $\hat{y}$ is a unit vector in the $y$-direction. The stream function $\psi(x,z,t)$ is defined by $u \! = \! \partial \psi/\partial z$ and $w \! =\!  - \partial \psi/\partial x.$ 

The no-slip boundary condition yields $\psi = \partial \psi/\partial z \!=\! 0$ at the top and bottom walls $z\!=\!0,1$ and $\psi \!=\! \partial \psi/\partial x \!=\! 0$ at the sidewalls $x\!=\!0,\Gamma_x$. The no-flux boundary condition yields $\partial c / \partial z = 0$ at $z=0,1$ and $\partial c / \partial x \!=\! 0$ at $x\!=\!0,\Gamma$.

The vorticity and the stream function are related by the Poisson equation
\begin{equation}
\omega = - \left( \frac{\partial^2 \psi}{\partial x^2} + \frac{\partial^2 \psi}{\partial z^2} \right). \label{eq:omega}
\end{equation}
The boundary conditions for $\omega$ are computed using $\psi$ and Eq.~(\ref{eq:omega}) evaluated at the boundaries.  The initial conditions are no fluid motion such that $\psi \!=\! \omega \!=\! 0$ everywhere with a concentration profile given by $c(x,z,t\!=\!0)\!=\!e^{-(\xi/\text{Le})^{1/2} x}$.

We expand $\psi$, $\omega$ and $c$ as a power series using $\text{Ra}_s$ as the small parameter
\begin{eqnarray}
\psi(x,z,t) &=& \psi_0(x,z,t) + \text{Ra}_s \psi_1(x,z,t) + \ldots \\ 
c(x,z,t) &=& c_0(x,z,t) + \text{Ra}_s c_1(x,z,t) + \ldots \\
\omega(x,z,t) &=& \omega_0(x,z,t) + \text{Ra}_s \omega_1(x,z,t) + \ldots
\end{eqnarray}
These expansions are inserted into Eq.~(\ref{eq:omegapsi})-(\ref{eq:omega}) and the equations are solved numerically for $\psi_i$, $c_i$, and $\omega_i$ at each order $i$ of $\text{Ra}_s^i$ using the appropriate boundary and initial conditions.

At $\mathcal{O}(0)$, Eq.~(\ref{eq:omegapsi}) yields the trivial solution $\omega_0 \!=\! \psi_0 \!=\! 0$ indicating no fluid motion $u\!=\!w\!=\!0$ as expected in the absence of solutal feedback. In this case, Eq.~(\ref{eq:cpsi}) becomes the reaction-diffusion equation for $c_0$,
\begin{equation}
\frac{\partial c_0}{\partial t} = \text{Le}\left(\frac{\partial^2 c_0}{\partial x^2} + \frac{\partial^2 c_0}{\partial z^2} \right) + \xi c_0(1-c_0). \label{eq:c0}
\end{equation}
The boundary conditions are $\partial c_0/\partial x\!=\!0$ at $x\!=\!0,\Gamma$ and $\partial c_0/\partial z=0$ at $z=0,1$. The initial condition is $c_0(x,z,t\!=\!0)\!=\!e^{-(\xi/\text{Le})^{1/2} x}$. For our boundary conditions and initial condition, $c_0$ is independent of $z$ such that $c_0(x,t)$ and, as a result,  Eq.~(\ref{eq:c0}) reduces further to the one dimensional reaction diffusion equation
\begin{equation}
\frac{\partial c_0}{\partial t} = \text{Le} \frac{\partial^2 c_0}{\partial x^2} + \xi c_0(1-c_0). \label{eq:c0-simple}
\end{equation}
This yields a vertically oriented front traveling with a front velocity of $v_0 = 2\sqrt{\text{Le} \xi}$. For the FKPP nonlinearity there is not a general explicit analytical solution for $c_0(x,z,t)$ (\emph{c.f.}~\cite{ablowitz:1979,brazhnik:1999}) and Eq.~(\ref{eq:c0-simple}) must be solved numerically.

\begin{figure}[tbh]
   \begin{center}
   \includegraphics[width=1.65in]{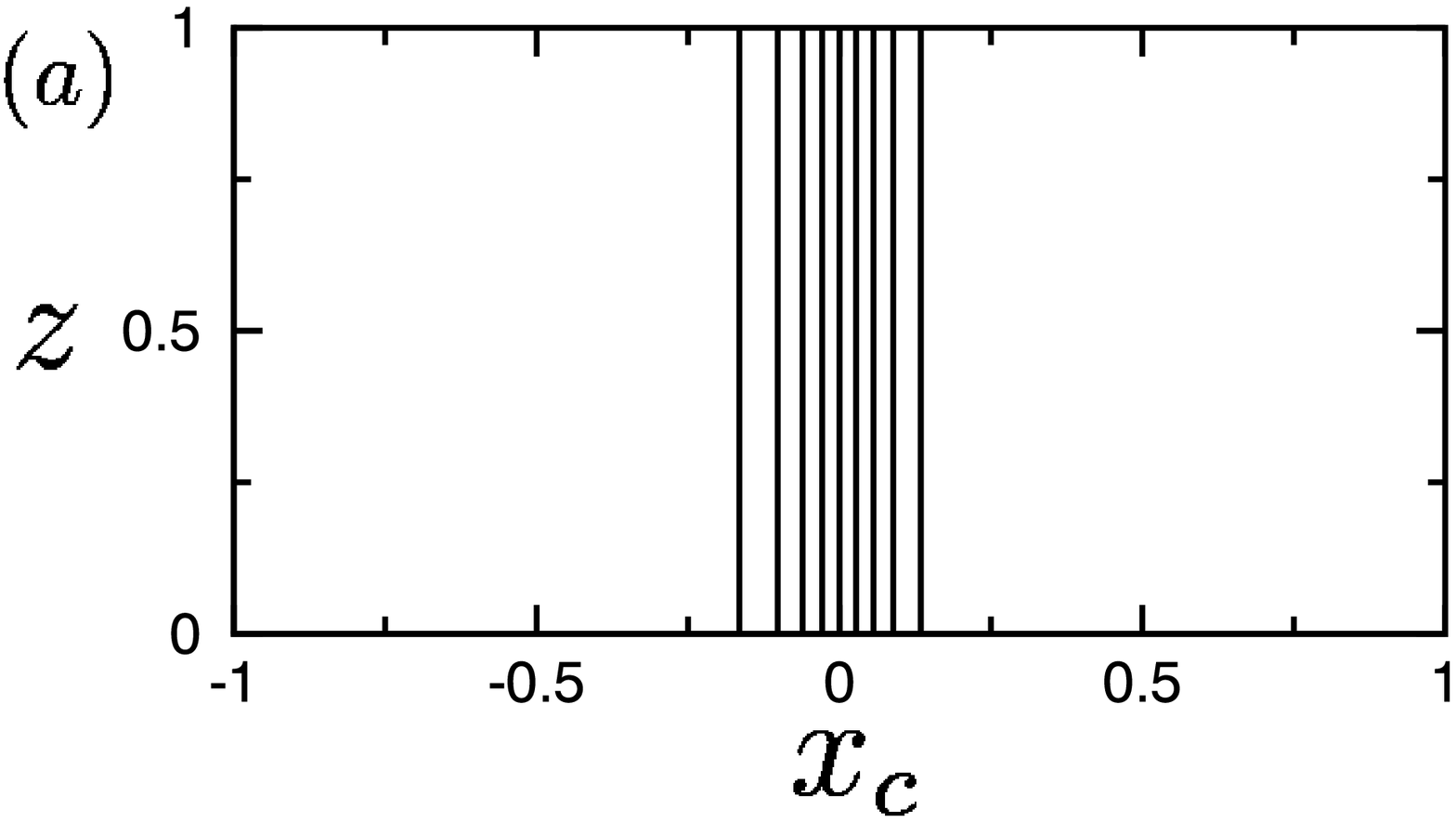} 
   \includegraphics[width=1.65in]{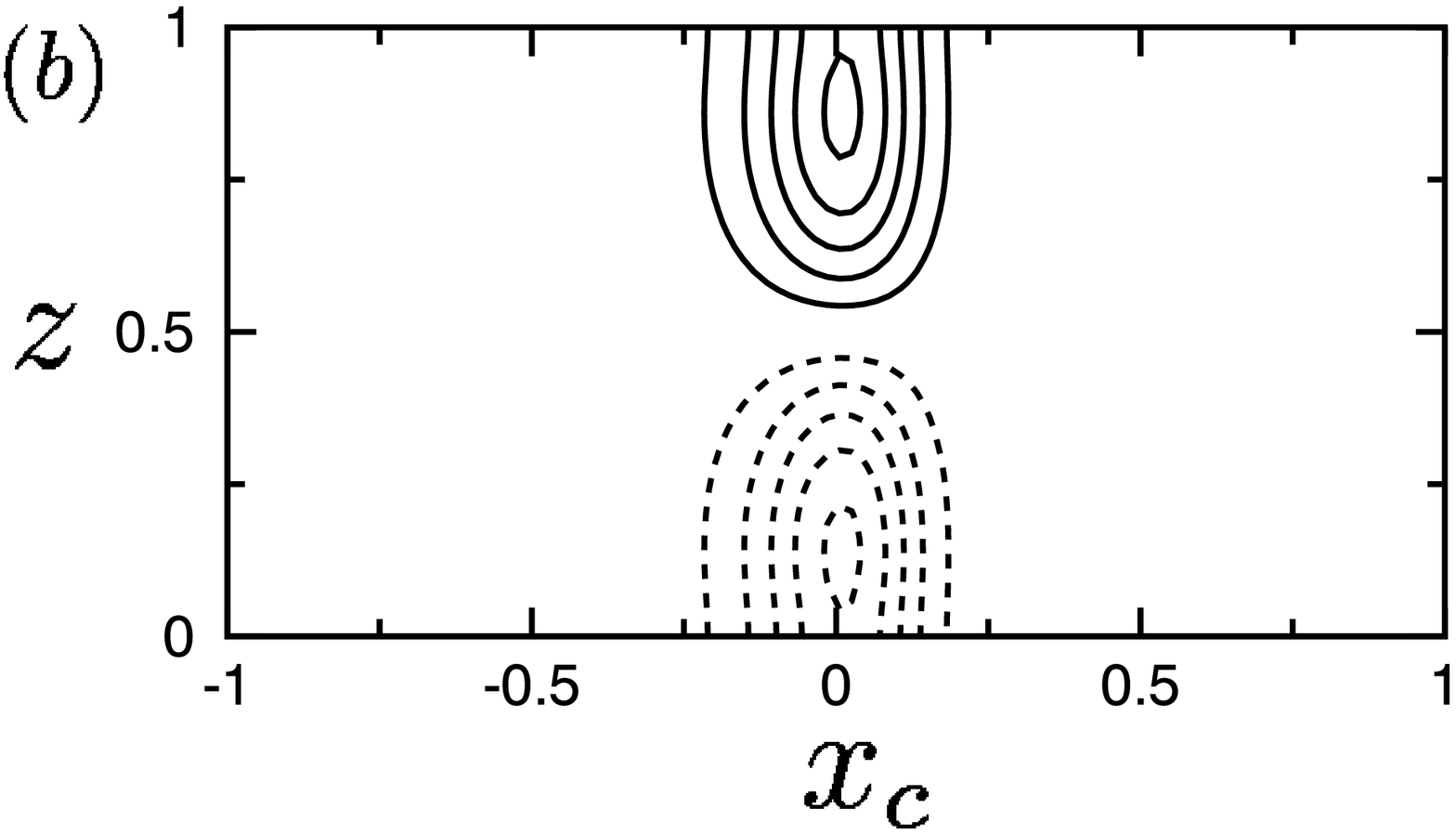}\\
   \includegraphics[width=1.65in]{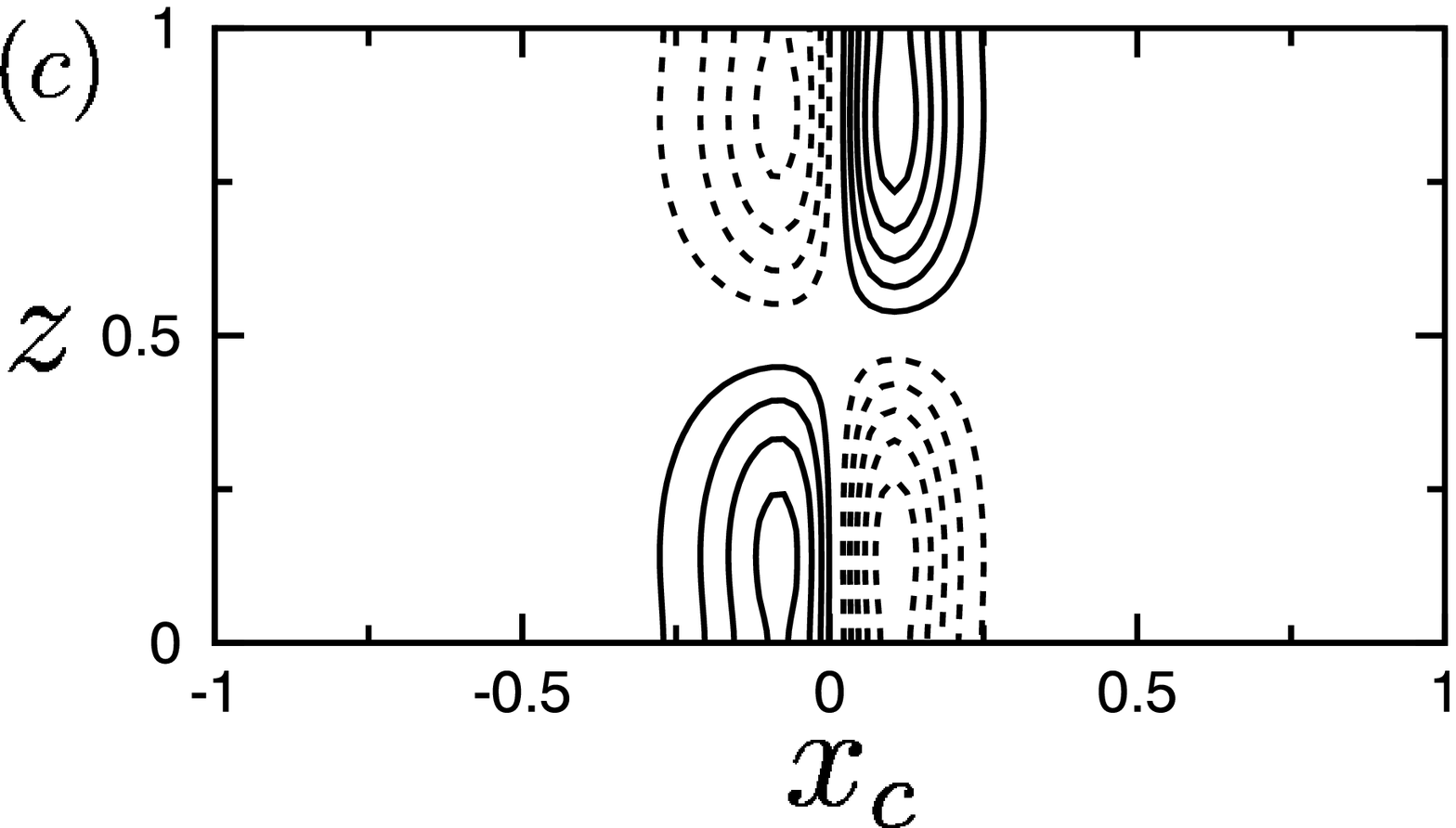}
   \includegraphics[width=1.65in]{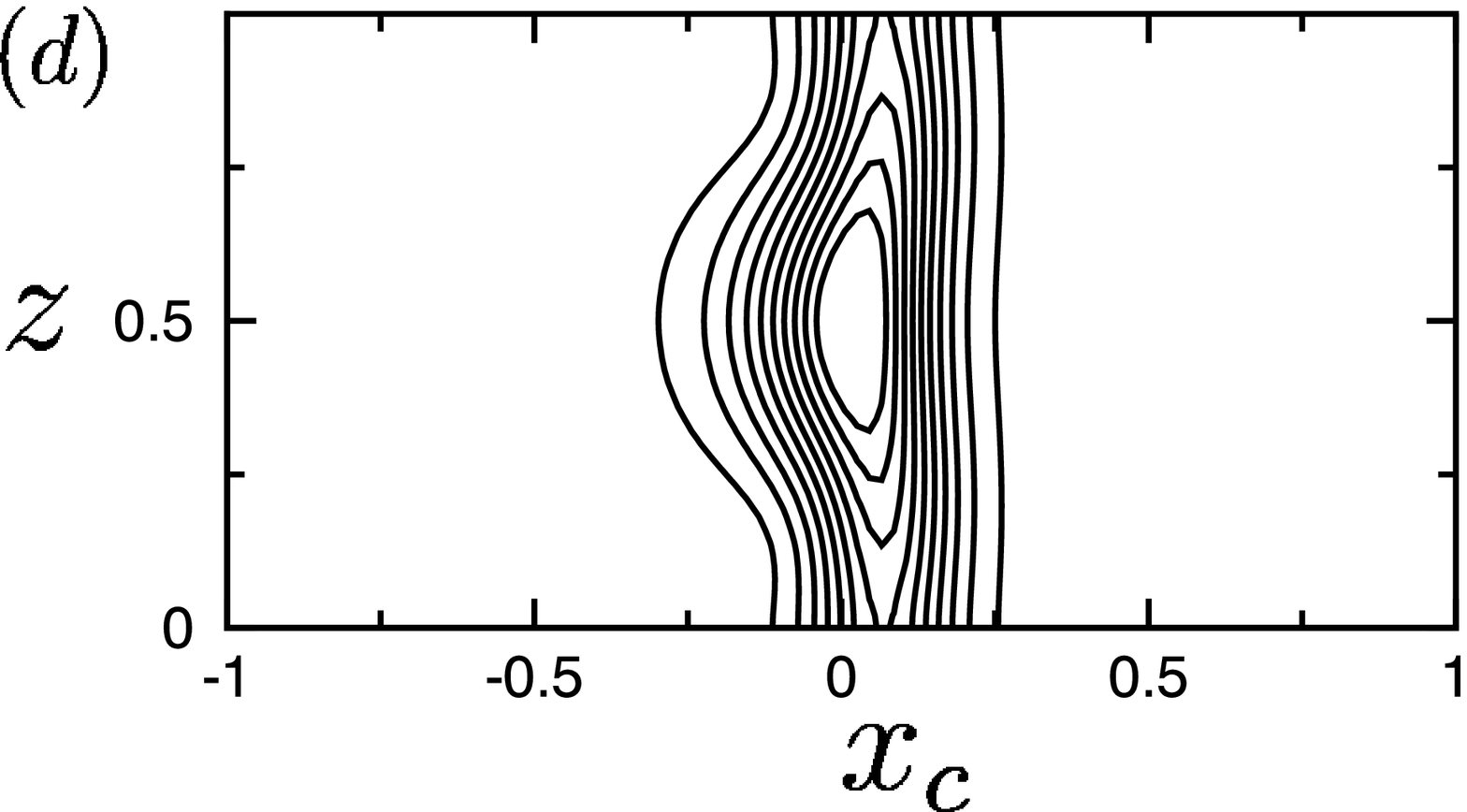}
   \caption{The spatial variation of (a)~$c_0$, (b)~$c_1$, (c)~$\frac{\partial c_1}{\partial t}$, and (d)~$c_2$ for a front at its asymptotic state for $\text{Ra}_s \! \ll \! 1$. Isocontours of the concentration are shown as solid (dashed) lines for positive (negative) values. The $x$ axis is scaled such that the isocontour $c_0(x,t) \!= \!1/2$ is located at $x_c\!=\!0$. (a)~The isocontours of $c_0$ are shown between 0.9 (left) and 0.1 (right) with a contour spacing of 0.1. $c_0$ is asymmetric about $x_c$.  (b)~The isocontours of $c_1$ are antisymmetric about $z\!=\!1/2$. Solid and dashed lines are equally spaced contours in $0.014 \! \le \! c_1 \! \le \! 0.07$ and $-0.07 \! \le \! c_1 \! \le \! -0.014$, respectively. The closed contour near the top (bottom) is the largest (smallest) value and the magnitude decreases (increases) monotonically moving outward. (c)~Isocontours of $\frac{\partial c_1}{\partial t}$ are antisymmetric about $z\!=\!1/2$. Solid lines are equally spaced contours in $0.05 \! \le \! \frac{\partial c_1}{\partial t} \! \le \! 0.25$. Dashed lines are equally spaced contours in $-0.05 \! \le \! \frac{\partial c_1}{\partial t} \! \le \! -0.25$. (d)~Equally spaced isocontours of $c_2$ between $0.001 \le c_1 \le 0.0145$. The largest value is located at the closed contour in the center and the magnitude decreases going outward. The curved front shape $c(x,z)$ that these variations in $c_0$, $c_1$ and $c_2$ yield for $\text{Ra}_s \! = \! 10^{-3}$ is shown by the blue curve in Fig.~\ref{fig:axial-flow-profiles}.} 
  \label{fig:c-pert}
  \end{center}
\end{figure}

The spatial variation of $c_0$ for a front at its asymptotic long-time state is shown in Fig.~\ref{fig:c-pert}(a). The solid lines are equally spaced isocontours of $c_0$ with a spacing of $\Delta c_0 \!=\! 0.1$  where the contour to the furthest left is $c_0\!=\!0.9$ and the contour to the furthest right is $c_0 \!=\! 0.1$. The axial position of the front is plotted using the coordinate $x_c$ where $x_c$ is the position relative to the location of the isocontour of $c_0=1/2$. Therefore, using this convention, $x_c=0$ is the location of the $c_0=1/2$ isocontour.  We highlight that $c_0(x)$ is asymmetric about $x_c=0$ which is evident by the variation of the spacing between the contour lines in Fig.~\ref{fig:c-pert}(a).  The mixing length $\bar{L}_s$ at $\mathcal{O}(0)$ is the axial distance between the 0.01 and 0.99 contours which yields a value of $L_0 = 0.608$.
%%%%%%%%%%%%%%%%%%%%%%%%%%%%%%%%%%%%%%%%%%%%
%%%%%%%%%%%%%%%%%%%%%%%%%%%%%%%%%%%%%%%%%%%%

The equations at $\mathcal{O}(1)$ are,
\begin{equation}
\text{Pr}^{-1} \frac{\partial \omega_1}{\partial t} = \frac{\partial^2 \omega_1}{\partial x^2} + \frac{\partial^2 \omega_1}{\partial z^2} + \frac{\partial c_0}{\partial x}
\label{eq:omegapsi1}
\end{equation}
and
\begin{multline}
\frac{\partial c_1}{\partial t} + \frac{\partial \psi_1}{\partial z} \frac{\partial c_0}{\partial x} = \text{Le}\left(\frac{\partial^2 c_1}{\partial x^2} + \frac{\partial^2 c_1}{\partial z^2} \right) \\ + \xi c_1 (1-2c_0)
\label{eq:cpsi1}
\end{multline}
where the vorticity and stream function are related by
\begin{equation}
\omega_1 = - \left( \frac{\partial^2 \psi_1}{\partial x^2} + \frac{\partial^2 \psi_1}{\partial z^2} \right). \label{eq:omega1}
\end{equation}

\begin{figure}[tbh]
   \begin{center}
   \includegraphics[width=2.5in]{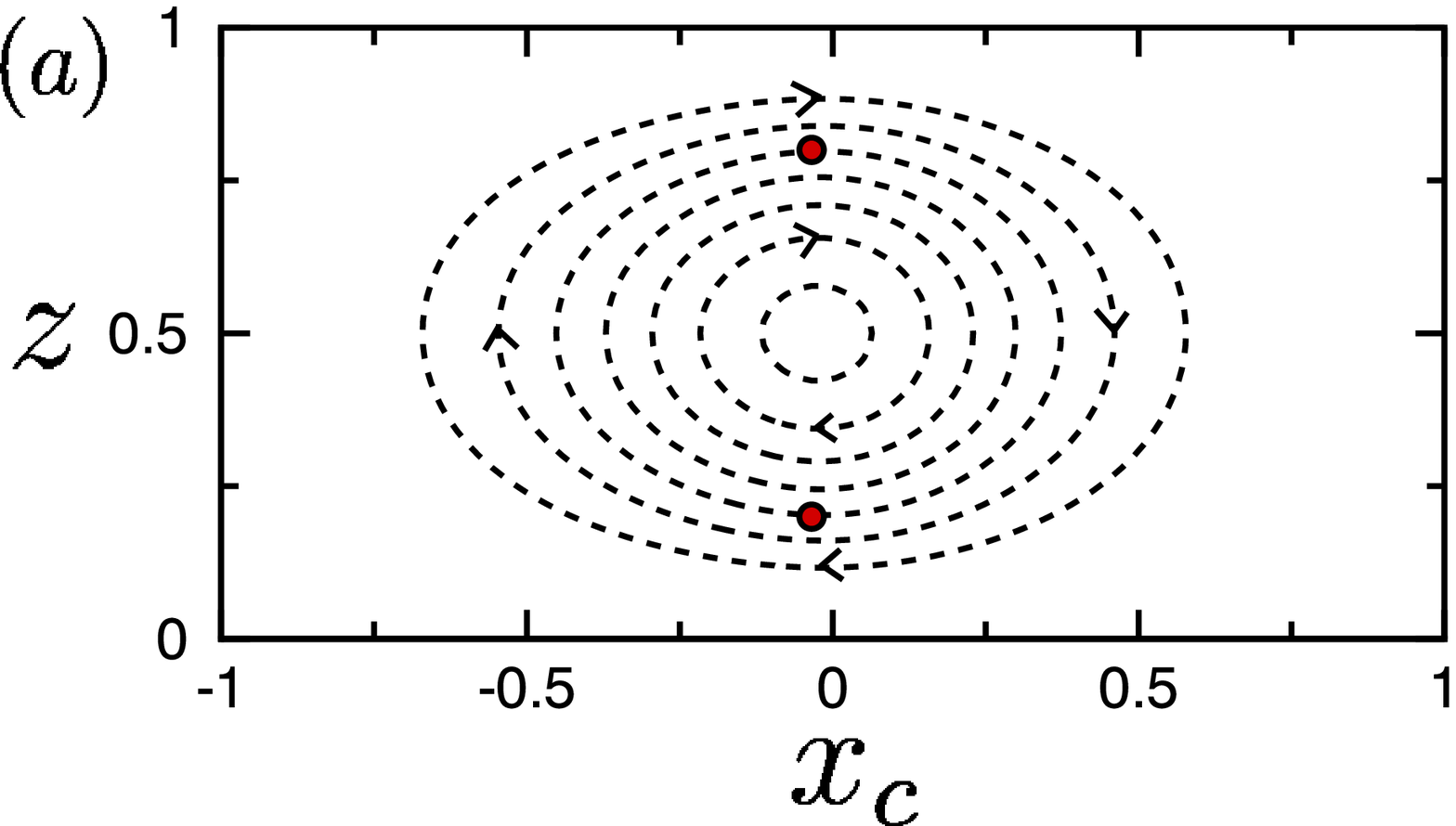}
   \includegraphics[width=2.5in]{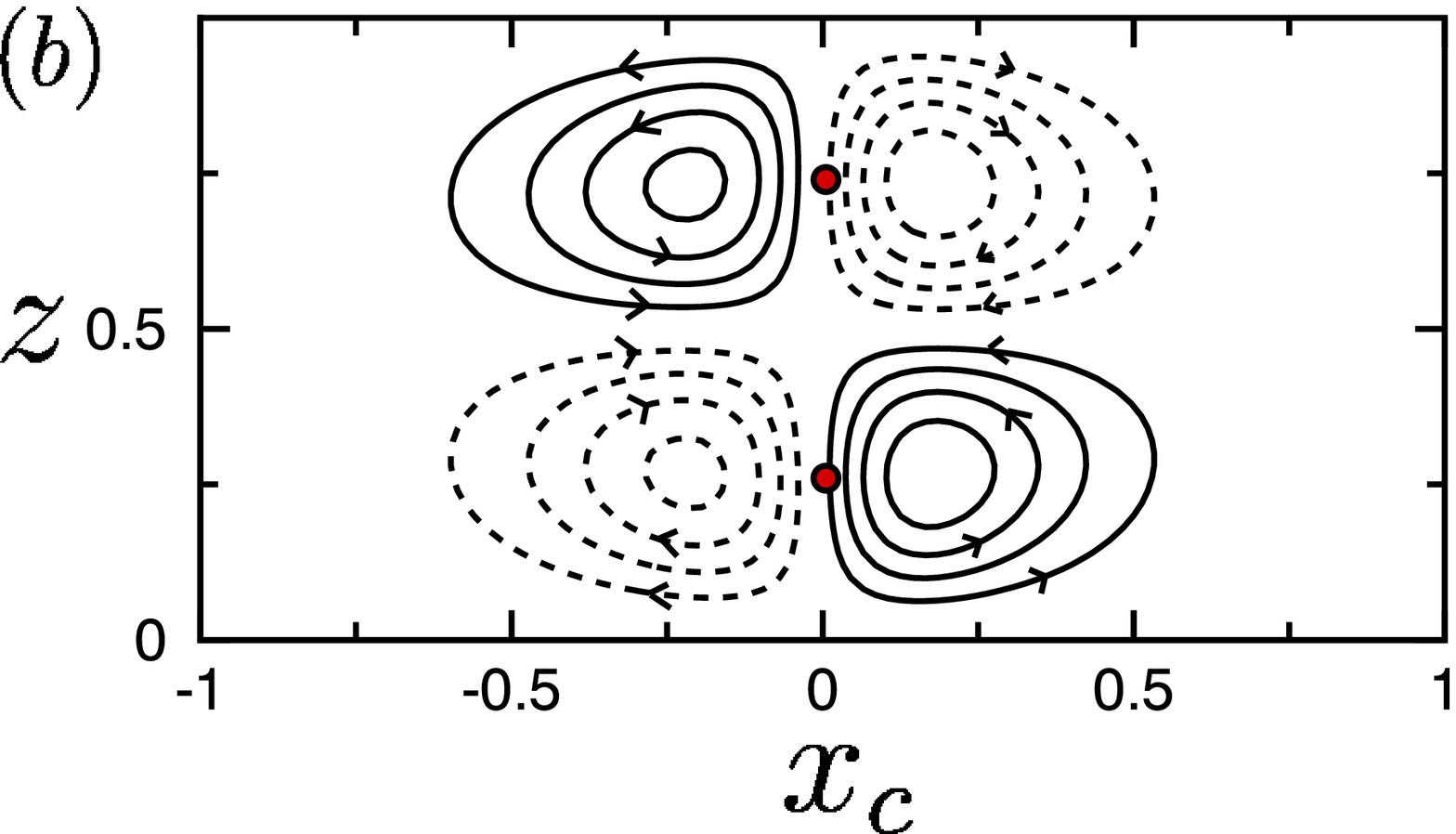}
   \caption{The spatial variation of (a)~$\psi_1(x,z)$ and (b)~$\psi_2(x,z)$ for a front at its asymptotic state for $\text{Ra}_s \ll 1$. Isocontours of the stream function are shown as solid (positive) and dashed (negative) lines and the arrows indicate the direction of fluid motion. The $x$ axis is scaled as in Fig.~\ref{fig:c-pert}. (a) $\psi_1$ is a vortical flow rotating clockwise. The circles (red) indicate the location of the maximum fluid velocity. Equally spaced isocontours are shown for $-3\times10^{-3} \!  \le \! \psi_1 \! \le \! -6 \times 10^{-4}$.  $\psi_1$ is largest at the center of the vortex and decreases with distance from the center. (b)~$\psi_2$ is a quadrupole of fluid flow.  Equally spaced isocontours are shown for $-2\times10^{-5}\! \le \! \psi_2 \! \le \! 2 \times 10^{-5}$ where the largest and smallest values are located at the centers of the vortex structures.} 
  \label{fig:psi-pert}
  \end{center}
\end{figure}

The vorticity $\omega_1(x,z,t)$ is nonzero and is driven by the spatial variation of $c_0(x,t)$ in the $x$-direction as indicated by Eq.~(\ref{eq:omegapsi1}). This results in a clockwise vortex of fluid motion as shown by the streamlines in Fig.~\ref{fig:psi-pert}(a). The center of this vortex occurs at $x_c < 0$ indicating that it is slightly to the left of the axial location of the $c_0=1/2$ isocontour line.

Therefore, the leading order contribution to the fluid motion is at $\mathcal{O}(1)$.  The magnitude of the maximum contribution to the fluid velocity at $\mathcal{O}(1)$, which we will refer to as $u_{1,\text{max}}$, is the axial velocity that occurs near the top and bottom of the domain. The location of $u_{1,\text{max}}$ is shown by the two circles (red) in Fig.~\ref{fig:psi-pert}(a) and has a value of $u_{1,\text{max}} = 9.6 \times 10^{-3}$.

Using our definition of the characteristic velocity $U$ as the maximum fluid velocity, we can represent $U$ to $\mathcal{O}(1)$ as $U = u_{1,\text{max}} \text{Ra}_s$. This yields $U = 9.6 \times 10^{-3} \text{Ra}_s$  which is indicated by the solid line in Fig.~\ref{fig:fluid-and-front-velocity-no-convection}(b). The agreement is excellent with the results from the full numerical simulations shown as the circles (blue). Therefore, the linear scaling of the fluid velocity is due to the axial variation of the concentration of the bare front which drives the vorticity field.

Equation~(\ref{eq:cpsi1}) indicates that the concentration $c$, through the variations of $c_1$, will now be altered from the vertical stripe structure of $c_0$ by the vortical flow field generated by $\psi_1$. The spatial variation of $c_1(x,z)$ is shown in Fig.~\ref{fig:c-pert}(b). $c_1$ is asymmetric in the $x$-direction about $x_c=0$ and is antisymmetric about the horizontal midplane $z=1/2$. The antisymmetry about the midplane has several important implications.

The variations of $c_1(x,z,t)$ cause the front to tilt toward the right and to develop some curvature at $\mathcal{O}(1)$. However, the mixing length is computed using the vertical average of the concentration field given by Eq.~(\ref{eq:cave}). Since $c_1(x,z)$ is antisymmetric about $z=1/2$, the $z$-average of $c_1$ will vanish and, as a result, the spatial variation of $c_1$ will not affect the value of the mixing length $\bar{L}_s$.

Similarly, using symmetry arguments, the variation of the front velocity $\bar{v}_f$ is also unaffected by the variations of $c_1$.  The $\mathcal{O}(1)$ contributions to the front velocity depend upon the $z$-average of $\frac{\partial c_1}{\partial t}$ as indicated by Eq.~(\ref{eq:vf}). The spatial variation of $\frac{\partial c_1}{\partial t}$ is shown in Fig.~\ref{fig:c-pert}(c) illustrating that it is antisymmetric about the horizontal midplane. As a result, the $z$-average of $\frac{\partial c_1}{\partial t}$ will vanish and there will not be an $\mathcal{O}(1)$ contribution to the front velocity. 
%%%%%%%%%%%%%%%%%%%%%%%%%%%%%%%%%%%%%%%%%%%%
%%%%%%%%%%%%%%%%%%%%%%%%%%%%%%%%%%%%%%%%%%%%

At $\mathcal{O}(2)$ the equations are
\begin{multline}
\text{Pr}^{-1} \left( \frac{\partial \omega_2}{\partial t}  + \frac{\partial \psi_1}{\partial z} \frac{\partial \omega_1}{\partial x} - \frac{\partial \psi_1}{\partial x} \frac{\partial \omega_1}{\partial z} \right) = \\ \frac{\partial^2 \omega_2}{\partial x^2} + \frac{\partial^2 \omega_2}{\partial z^2} + \frac{\partial c_1}{\partial x},
\label{eq:omegapsi2}
\end{multline}
and
\begin{multline}
\frac{\partial c_2}{\partial t} +  \frac{\partial \psi_2}{\partial z} \frac{\partial c_0}{\partial x} + \frac{\partial \psi_1}{\partial z} \frac{\partial c_1}{\partial x} - \frac{\partial \psi_1}{\partial x} \frac{\partial c_1}{\partial z} = \\ \text{Le}\left(\frac{\partial^2 c_2}{\partial x^2} + \frac{\partial^2 c_2}{\partial z^2} \right) + \xi (c_2 (1-2c_0) - {c_1}^2) \label{eq:cpsi2}
\end{multline}
with the relevant Poisson equation that is similar to Eq.~(\ref{eq:omega1}) but is in terms of $\omega_2$ and $\psi_2$.
In writing Eqs.~(\ref{eq:cpsi1}) and~(\ref{eq:cpsi2}) we have used the fact that $c_0$ is not a function of $z$ to simplify the expressions. The spatial variation of $c_2(x,z)$ and $\psi_2(x,z)$ are shown in Figs.~\ref{fig:c-pert}(d) and~\ref{fig:psi-pert}(b), respectively.

The stream function $\psi_2$ is a quadrupole of fluid motion as indicated by the streamlines in Fig.~\ref{fig:psi-pert}(b). From the streamlines it is evident that $\psi_2$ is asymmetric about its center in the $x$-direction and it is antisymmetric about the midplane $\!z=\!1/2$. The center of $\psi_2$ aligns with the center of $\psi_1$ which is slightly to the left of $c_0\!=\!1/2$ contour. The largest magnitude of the fluid velocity at $\mathcal{O}(2)$ occurs in the lobes of the closed contours located at $x_c \!>\! 0$ and are indicated by the red circles.

The concentration field $c_2$ is asymmetric in both the $x$ and $z$ directions. In particular, $z$ averages of $c_2$ and $\frac{\partial c_2}{\partial t}$ are nonzero and lead to contributions to $\bar{L}_s$ and $\bar{v}_f$. To $\mathcal{O}(2)$ this yields the following expression for the mixing length $(\bar{L}_s\!-\!L_0)/L_0\!=\!8.55 \times 10^{-3} \text{Ra}_s^2$ which is indicated by the solid line in Fig.~\ref{fig:solutal-roll}(c). Similarly, the front velocity to $\mathcal{O}(2)$ is given by $(\bar{v}_f - v_0)/v_0 = 1.635 \times 10^{-4}  \text{Ra}_s^2$ which is indicated by the solid line in Fig.~\ref{fig:fluid-and-front-velocity-no-convection}(d). The agreement between the perturbation analysis and the full numerical simulations is excellent. Overall, these results indicate that the absence of $\mathcal{O}(1)$ contributions to $\bar{L}_s$ and $\bar{v}_f$ is due to the antisymmetry of $c_1(t)$ and $\partial c_1 /\partial t$ about the horizontal midplane which leads to the quadratic scaling where this symmetry is broken.

\subsection{A Front with Solutal Feedback Propagating through a Convective Flow Field}

We next discuss how solutal feedback affects a front that propagates through a cellular convective flow field. In order to establish a convective flow field we used a thermal Rayleigh number of $\text{Ra}_T = 3000$. We first ran a long-time simulation of the flow field at this value of $\text{Ra}_T$ to establish a steady field of counter-rotating convection rolls over the entire domain. We accomplished this by using a hot-wall boundary condition at the sidewalls of the domain such that $T(x\!=\!0,z)\!=\!T(x\!=\!\Gamma,z)\!=\!1$. These boundary conditions drive an upflow near the sidewalls which initiates the formation of convection rolls near the walls that eventually fill the entire domain. For our numerical simulation using $\Gamma = 30$ this resulted in 30 convection rolls which yields an average roll width of unity.

In our simulations this yielded a characteristic velocity of the convective fluid motion, in the absence of solutal feedback, of $\bar{U}_c \!=\! 10.81$. As a result, the ratio of the convective fluid velocity time scale to the reaction time scale yields a Damk\"ohler number of $\text{Da} \!=\! \xi/\bar{U}_c \!\approx\! 1$ which indicates that the convection and reaction time scales are comparable. Furthermore, the ratio of fluid convection to mass diffusion yields a P\'eclet number of $\text{Pe}\!=\!\bar{U}_c/\text{Le} \!\approx\! 1000$ indicating that the thermal convection driven fluid velocity is significant. We have not explored the fronts for a broader range of convective flows in the presence of solutal feedback and this is a topic of future interest.

\begin{figure}[tbh]
   \begin{center}
   \includegraphics[width=3.4in]{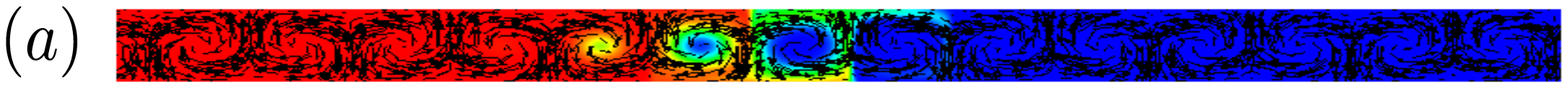} \\ 
   \includegraphics[width=3.4in]{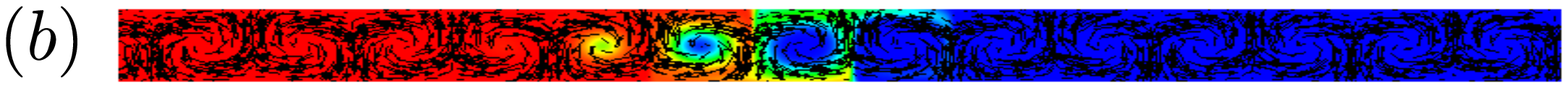} \\ 
   \includegraphics[width=3.4in]{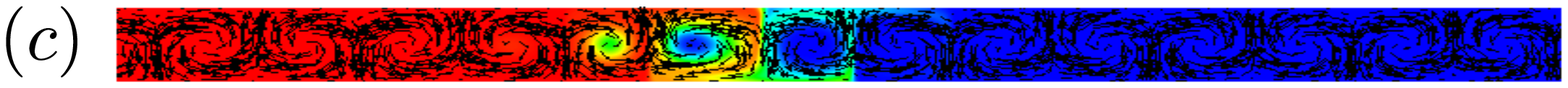} \\ 
   \includegraphics[width=3.4in]{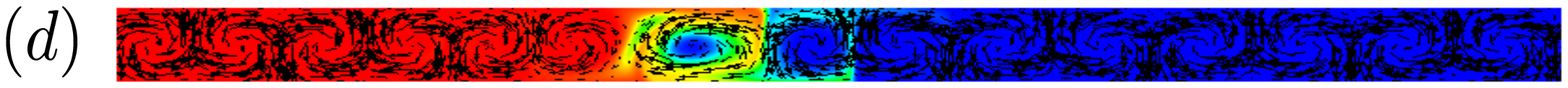} \\ 
   \includegraphics[width=3.4in]{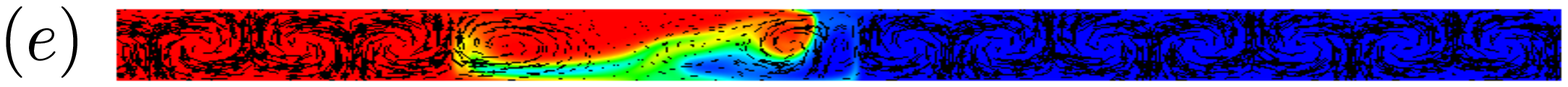} \\ 
   \includegraphics[width=3.4in]{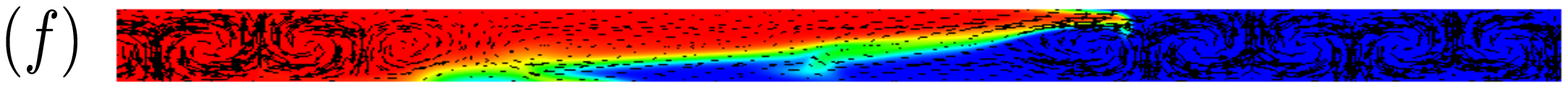} \\ 
   \includegraphics[width=3.4in]{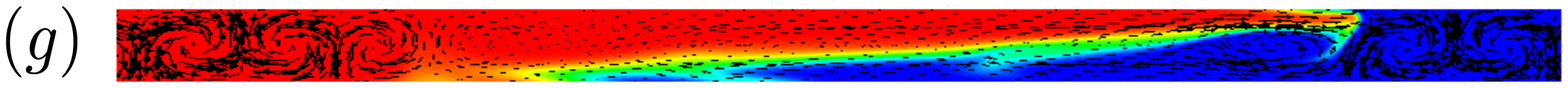} \\ 
   \vspace{0.2cm}
   \includegraphics[width=0.75in]{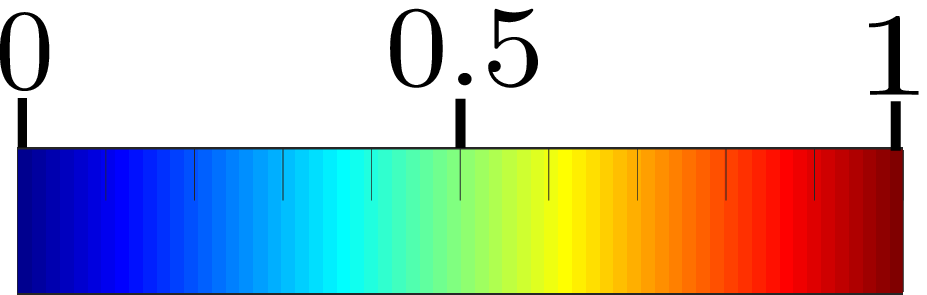} 
   \caption{Fronts propagating through a convective flow field with solutal feedback. $\text{Ra}_T \! = \! 3000$ and each panel is for a different value of $\text{Ra}_s$ at time $t\!=\!3$. Color shows $c$ where red is products $(c \!=\! 1)$ and blue is reactants $(c\!=\!0)$. The black arrows are of the fluid velocity $\vec{u}$. (a)-(g):~$\text{Ra}_s \!=\! \{0,100,500,700,1000,2000,3000\}$, respectively. A zoomed in view is shown where $3\!\le \!x \le \! 17$.}
    \label{fig:fronts-with-convection}
    \end{center}
\end{figure}

Images of the flow fields and propagating fronts are shown in Fig.~\ref{fig:fronts-with-convection}. Color contours are of the concentration $c(x,z,t)$ using our typical convention where red is products and blue is reactants. The black arrows are fluid velocity vectors $\vec{u}$ which make visible the chain of counter-rotating convection rolls that have resulted from the convective instability.  The front has been initiated at the left wall and is propagating to the right. All fronts are shown at a time $t\!=\!3$ after the front initiation and only a portion of the domain is shown in order to visualize the flow field and front features.
%%%%%%%%%%%%%%%%%%%%%%%%%%%%%%%%%%%%%%%%%%%%
%%%%%%%%%%%%%%%%%%%%%%%%%%%%%%%%%%%%%%%%%%%%

Figure~\ref{fig:fronts-with-convection}(a) shows a front for $\text{Ra}_s\!=\!0$ where there is no solutal feedback which results in an unchanging flow field as shown. In addition, it is clear that the front dynamics are affected by the flow field which causes it to spiral toward the cores of the convection rolls while propagating toward the right.

Figure~\ref{fig:fronts-with-convection}(b)-(g) shows results for $\text{Ra}_s > 0$ where there is a complex interplay between the thermal convection and the solutal feedback caused by the reacting front. For small values of $\text{Ra}_s$, the solutally induced convection roll is weak compared to the convective rolls. As a result, panels (a) and (b) of Fig.~\ref{fig:fronts-with-convection} are quite similar. However, as $\text{Ra}_s$ increases the strength of the solutal convection roll increases and its interactions with the convection rolls causes distortions in the flow field near the front as shown in Fig.~\ref{fig:fronts-with-convection}(c)-(d). For further increases in $\text{Ra}_s$, the solutal convection roll dominates the thermal convection rolls as shown in Fig.~\ref{fig:fronts-with-convection}(e)-(g). For large values of $\text{Ra}_s$, the solutal convection roll extends for many convection roll widths and annihilates the convective motion over the region spanned by the front. After the front passes through a location, the convection rolls reemerge  due to the convective instability. This is illustrated by the convection rolls to the left of the front in the region occupied by pure products. 

\begin{figure}[tbh]
   \begin{center}
   \includegraphics[width=2.75in]{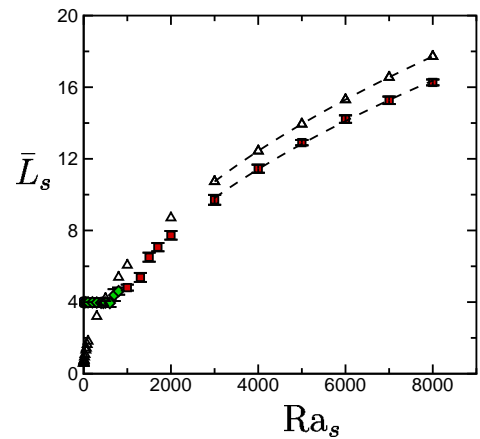} 
   \caption{The variation of the mixing length $\bar{L}_s$ for a front propagating through a convective flow field ($\text{Ra}_T\!=\!3000$) as a function of $\text{Ra}_s$ using our convention of circles (blue), diamonds (green), and squares (red) for low, intermediate, and large value of $\text{Ra}_s$, respectively. The mixing length for $\text{Ra}_T\!=\!0$ are included as the triangles for reference. The dashed lines indicate a scaling of $\bar{L}_s \propto \text{Ra}_s^{1/2}$.}
    \label{fig:mixing-length-with-convection}
    \end{center}
\end{figure}

Figure~\ref{fig:mixing-length-with-convection} shows the variation of the mixing length with $\text{Ra}_s$ for fronts propagating through convection rolls. The mixing length varies in time due to the interactions with the convection rolls. In Fig.~\ref{fig:mixing-length-with-convection} we show the time average value $\bar{L}_s$ using the filled symbols where the error bars indicate the standard deviation of the oscillations about the mean value.
%%%%%%%%%%%%%%%%%%%%%%%%%%%%%%%%%%%%%%%%%%%%
%%%%%%%%%%%%%%%%%%%%%%%%%%%%%%%%%%%%%%%%%%%%

For $\text{Ra}_s\!=\!0$ the value of the mixing length is $\bar{L}_s\!=\!4.0\!>\!L_0$ which represents the mixing length enhancement due to the convective flow field alone. A mixing length of 4 corresponds to two pairs of convection rolls since the width of a convection rolls is approximately unity.  From Fig.~\ref{fig:fronts-with-convection}(a) it is clear that the reaction zone spans approximately 4 convection rolls. The mixing length remains approximately at this value for all results where $\text{Ra}_s \! \lesssim \! 700$ which includes the circles (blue) and some of the diamonds (green) in Fig.~\ref{fig:mixing-length-with-convection}. As the solutal Rayleigh number increases $\text{Ra}_s \! \gtrsim \! 700$ the mixing length begins to grow as shown by the remaining diamonds (green) and the squares (red). For large values of $\text{Ra}_s$ the data scales as $\bar{L}_s \! \propto \! \text{Ra}_s^{1/2}$ as indicated by the dashed line.

\begin{figure}[tbh]
   \begin{center}
   \includegraphics[width=1.65in]{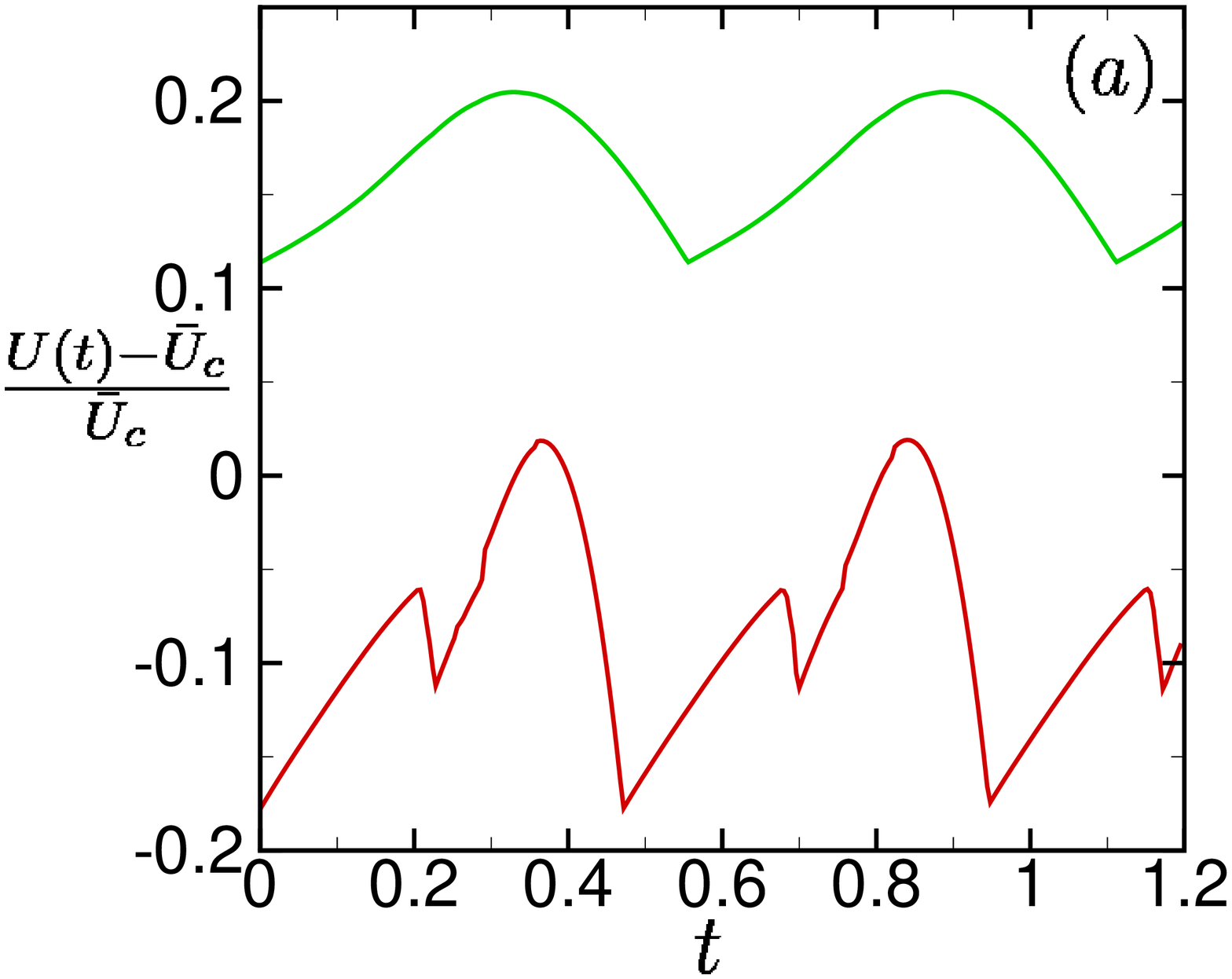} 
   \includegraphics[width=1.65in]{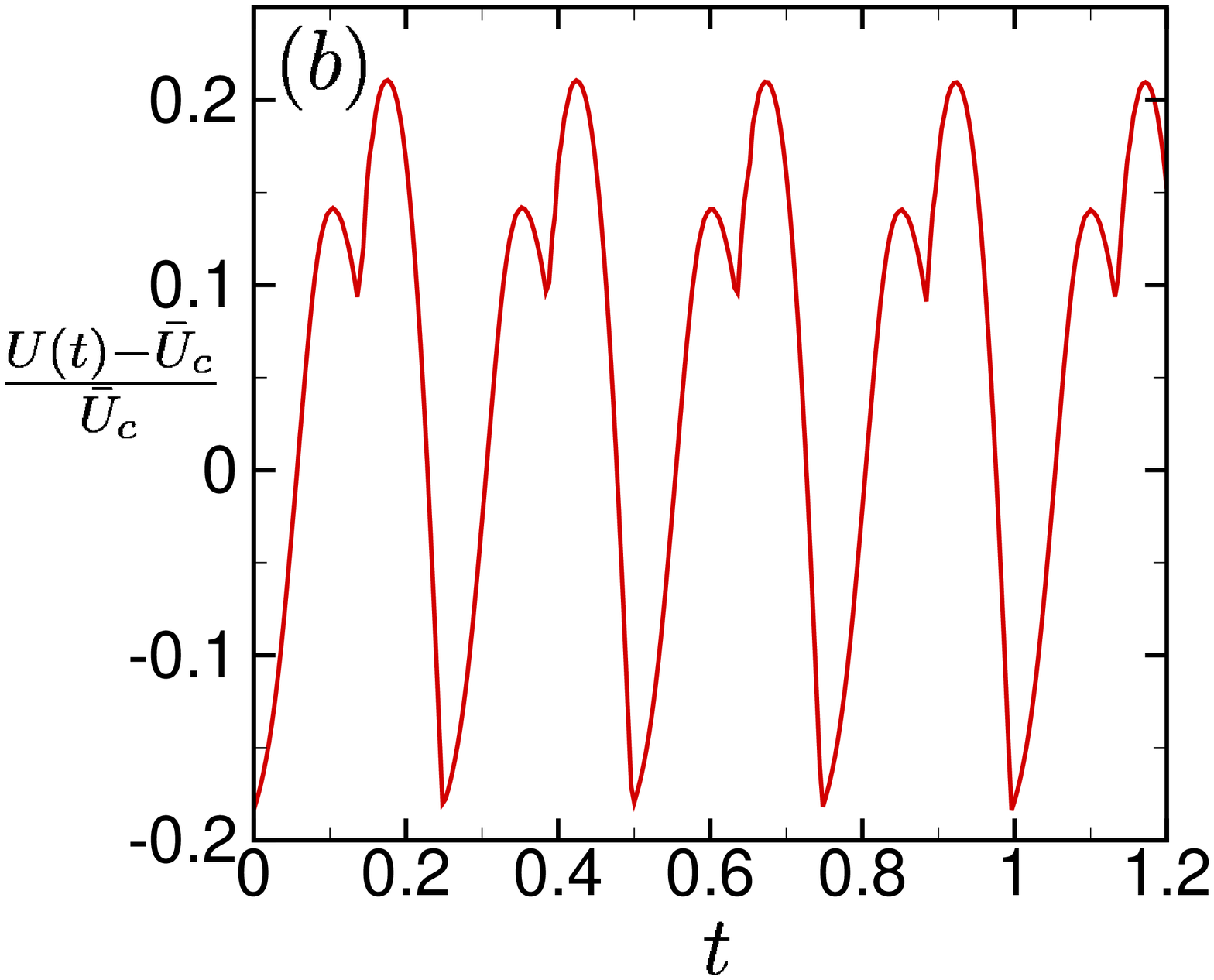}
   \includegraphics[width=1.8in]{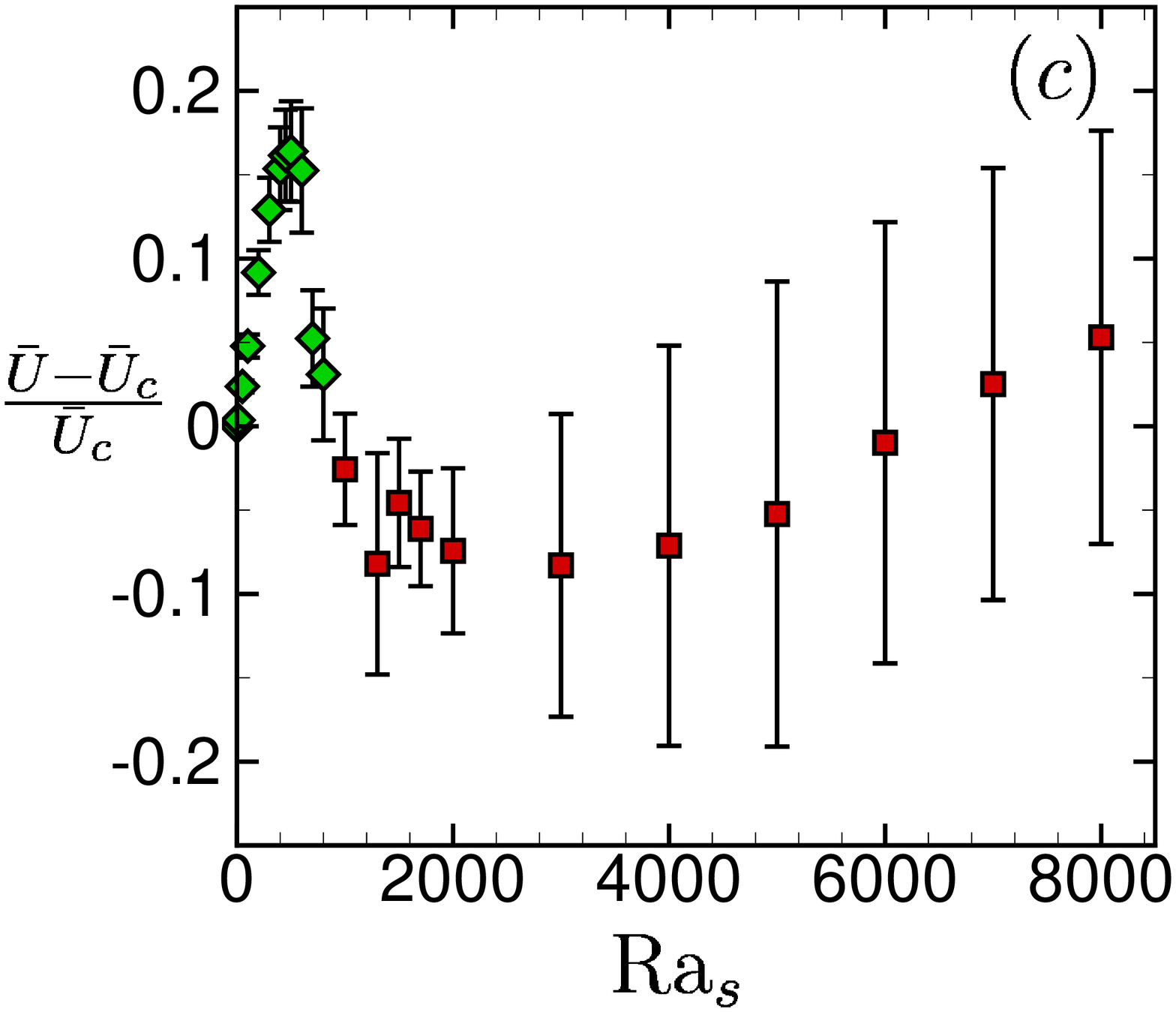}  
\caption{The variation of the scaled characteristic fluid velocity for a front propagating through a convective flow field with $\text{Ra}_T\!=\!3000$. The characteristic velocity of the background convective flow field in the absence of a front is $\bar{U}_c\!=\!10.81$.  (a)~The time variation of the normalized fluid velocity $(U(t)\!-\!\bar{U}_c)/\bar{U}_c$ for $\text{Ra}_s \!=\! 500$ (upper, green) and $\text{Ra}_s \!=\! 2000$ (lower, red) and in~(b) for $\text{Ra}_s \!=\! 8000$. In these plots time has been adjusted such that $t\!=\!0$ at the beginning of a period of the oscillatory dynamics for easier comparison.  (c)~The variation of the normalized mean value of the characteristic fluid velocity $(\bar{U}\!-\!\bar{U}_c)/\bar{U}_c$ with $\text{Ra}_s$ where the error bars represent the standard deviation of $U(t)$ about the mean value. Flow field images for these fronts are shown in Fig.~\ref{fig:fronts-with-convection}.}
    \label{fig:fluid-velocity-with-convection}
    \end{center}
\end{figure}

The mixing length results, in the absence of thermal convection ($\text{Ra}_T \!=\! 0$), are included as the triangles for comparison. The presence of the thermal convection causes $\bar{L}_s$ to be larger for very small $\text{Ra}_s$ and then smaller for larger values of $\text{Ra}_s$. The variation of the characteristic fluid velocity $U$ is shown in Fig.~\ref{fig:fluid-velocity-with-convection}.  For fonts propagating through convective flow fields we define the characteristic fluid velocity $U(t)$ as the maximum fluid velocity that occurs in the spatial region around the front that we have previously identified as the mixing length $L_s$.

In Fig.~\ref{fig:fluid-velocity-with-convection}(a)-(c) we show $U(t)$ for several representative examples which demonstrate the oscillatory fluid dynamics that occur due to the solutal feedback of the propagating front. Figure~\ref{fig:fluid-velocity-with-convection}(c) shows the time average of the characteristic fluid velocity $\bar{U}$ over a large range of $\text{Ra}_s$ where the error bars are the standard deviations about the mean value of the oscillations. The fluid velocity is scaled using the characteristic fluid velocity of the convective flow field in the absence of solutal feedback $\bar{U}_c$. When presented this way, a positive (negative) velocity indicates a characteristic velocity that is larger (smaller) than the background convective flow field.
%%%%%%%%%%%%%%%%%%%%%%%%%%%%%%%%%%%%%%%%%%%%
%%%%%%%%%%%%%%%%%%%%%%%%%%%%%%%%%%%%%%%%%%%%

The upper curve (green) of Fig.~\ref{fig:fluid-velocity-with-convection}(a) illustrates the periodic dynamics of $U(t)$ for $\text{Ra}_s\!=\!500$ which corresponds to the case where the peak occurs in Fig.~\ref{fig:fluid-velocity-with-convection}(c). For this case, $U(t)$ is greater than the characteristic velocity of the background convective flow for all time. This indicates that the solutal feedback is increasing the fluid velocity.  The characteristic fluid velocity rises and then falls periodically. The periodic oscillation is due to the counter-rotating convection rolls. The leading edge of the propagating front is near the upper wall for $\text{Ra}_s \!>\! 0$ as shown in Fig.~\ref{fig:fronts-with-convection}. When the front approaches the left side of a counter-clockwise convection roll, the directions of the front and the fluid velocity are opposing. This interaction results in a reduction in $U(t)$ and the troughs of the green curve occur at these times. When the front approaches the left side of clockwise convection roll, the front and convective velocity are cooperative and this results an in increase in $U(t)$ and the peak values of the green curve in Fig.~\ref{fig:fluid-velocity-with-convection}(a).

The convection rolls have a spatial wavelength of $\lambda \!\approx\! 2$ since two rolls of unity width are required for the convective flow field to repeat. Therefore, we can use $U(t)$ to provide an estimate of the front velocity as $v_f(t) \! \approx \! \lambda /t_p$ where $t_p$ is the duration required for $U(t)$ to repeat in Fig.~\ref{fig:fluid-velocity-with-convection}(a). For the green curve this yields $v_f \approx 2/0.56 \!=\! 3.57$. This is approximate since the solutal feedback will distort the convection rolls such that $\lambda$ may change significantly for large values of $\text{Ra}_s$.

The lower curve (red) of Fig.~\ref{fig:fluid-velocity-with-convection}(a) shows $U(t)$ for $\text{Ra}_s\!=\!2000$ which corresponds to the case where $\bar{U}$ is small in Fig.~\ref{fig:fluid-velocity-with-convection}(c). For this case, $U(t)$ is less than the convective fluid velocity except for a brief time near its peak. In this case, the interaction of the solutal feedback with the convection rolls results in a decrease in the fluid velocity on average. There are now two peaks in $U(t)$ within the periodic dynamics. These two peaks are again related to the spatial locations where the convection rolls are either favorable or opposing to the front motion.  It is clear that the red curve repeats over a shorter duration than the green curve which suggests that the front velocity is larger for this case. For this case we find $v_f \! \approx \! 2/0.47 \! = \! 4.26$ which is larger as expected.

Figure~\ref{fig:fluid-velocity-with-convection}(b) illustrates $U(t)$ for the large value of $\text{Ra}_s \!=\! 8000$. In this case,  the periodic dynamics contain two peaks as expected for the interaction of the front with the counter-rotating convection rolls. The maximum value is positive and the minimum value is negative and the front is clearly now much faster. An estimate of the front velocity gives $v_f \approx 2/0.25 \!=\! 8.0$.

Figure~\ref{fig:fluid-velocity-with-convection}(c) illustrates the trend that $\bar{U}$ initially increases and reaches a peak value near $\text{Ra}_s \! \approx \! 500$. For larger values of the solutal Rayleigh number, $\bar{U}$ decreases and reaches a minimum near $\text{Ra}_s \! \approx \! 3000$. Further increases of $\text{Ra}_s$ yields increasing values of $\bar{U}$ for the range of our calculations.

\begin{figure}[tbh]
   \begin{center}
   \includegraphics[width=1.65in]{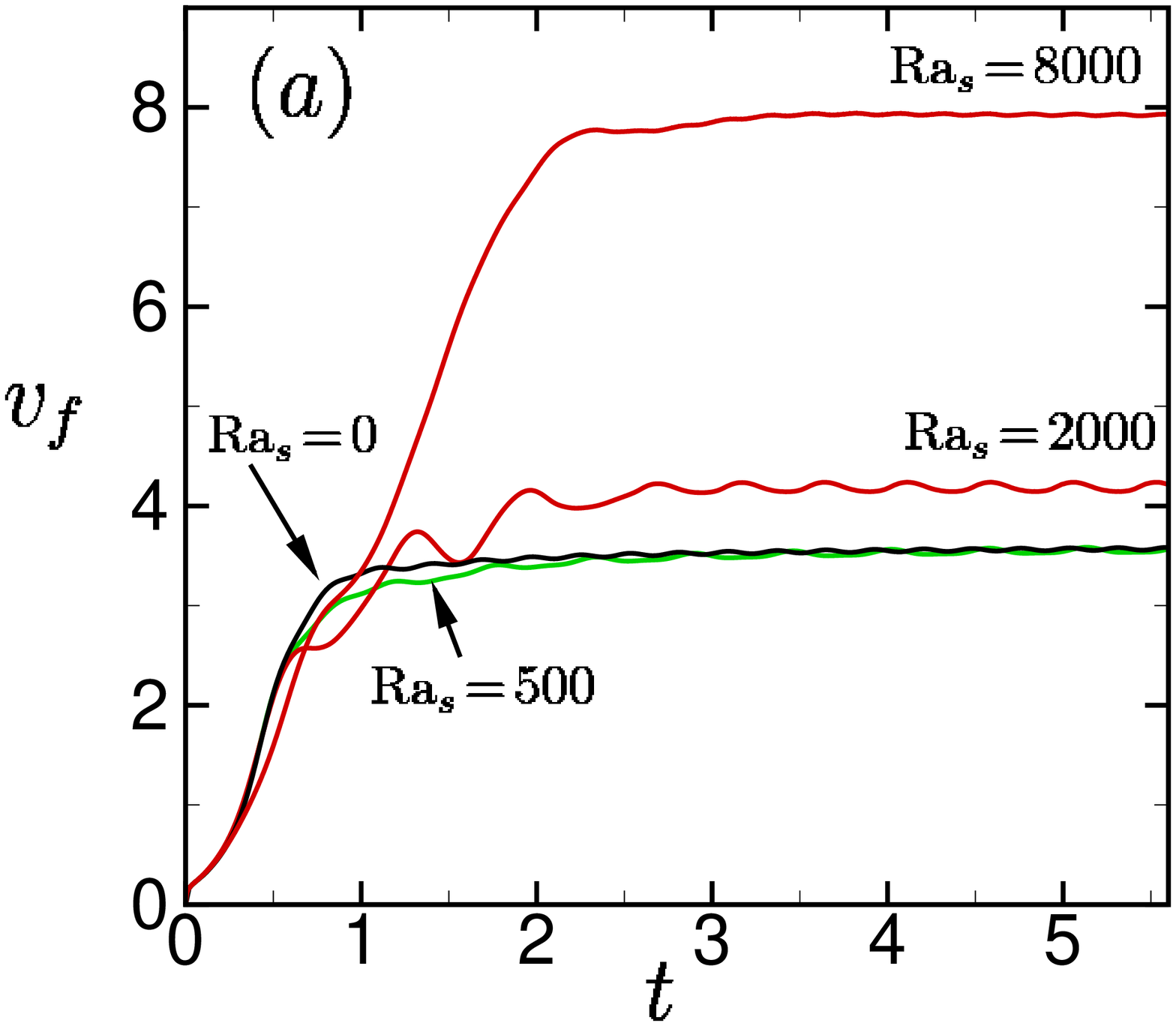} 
      \includegraphics[width=1.65in]{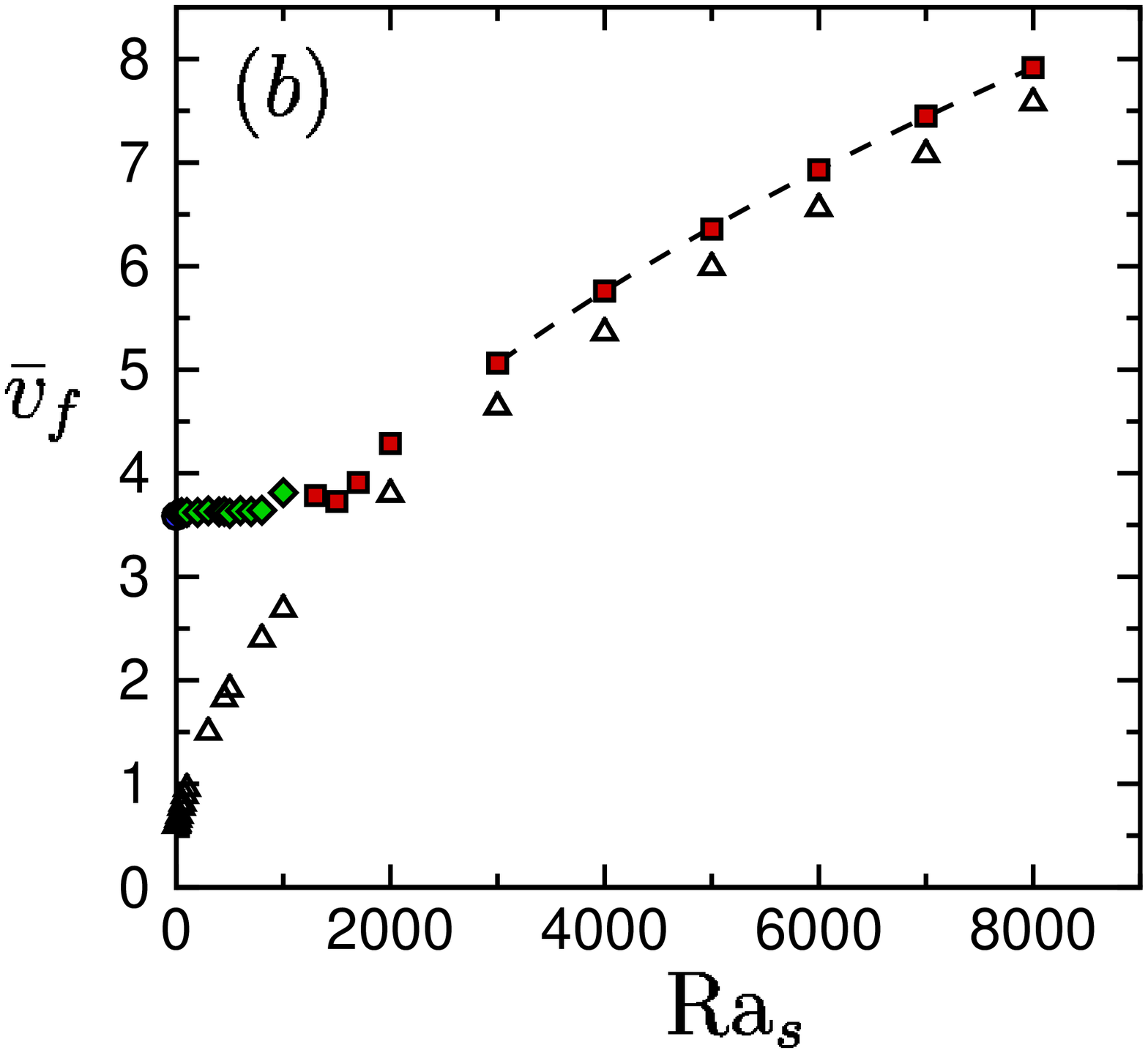} 
   \caption{The variation of the front velocity when propagating through convection rolls with $\text{Ra}_T \!=\! 3000$. (a)~The variation of the front velocity $v_f(t)$ with time $t$ for different values of $\text{Ra}_s$ where $\text{Ra}_s\!=\!0$ (black), $\text{Ra}_s\!=\!500$ (green), $\text{Ra}_s\!=\!2000$ (lower red) and $\text{Ra}_s\!=\!8000$ (upper red). (b)~The asymptotic front velocity $\bar{v}_f$ as a function of $\text{Ra}_s$. The front velocity when $\text{Ra}_s\!=\!0$ is $\bar{v}_f\!=\!3.59$. The dashed line represents a scaling of $\text{Ra}_s^{1/2}$.  Flow field images corresponding to these results are shown in Fig.~\ref{fig:fronts-with-convection}. The open triangles are the results for the front velocity in absence of convection from Fig.~\ref{fig:fluid-and-front-velocity-no-convection}(c) and are included here for comparison.}
    \label{fig:front-velocity-with-convection}
    \end{center}
\end{figure}

The variation of the front velocity is shown in Fig.~\ref{fig:front-velocity-with-convection}. Figure~\ref{fig:front-velocity-with-convection}~(a) shows $v_f(t)$ for several illustrative examples. The black curve is the front velocity for $\text{Ra}_s \!=\! 0$ and is the front velocity in the absence of solutal feedback.  Small oscillations are evident due to the convecting of the front by the fluid motion. The green curve shows $v_f(t)$ for $\text{Ra}_s=500$ which is very similar to $v_f(t)$ in the absence of solutal feedback. It is interesting to point out that the characteristic fluid velocity has a peak value at this value of $\text{Ra}_s$ as shown in Fig.~\ref{fig:fluid-velocity-with-convection}(c). The lower red curve shows $v_f(t)$ for $\text{Ra}_s \!=\! 2000$ which yields clear  temporal oscillations. Lastly, the upper red curve shows results for $\text{Ra}_s \!=\! 8000$.
%%%%%%%%%%%%%%%%%%%%%%%%%%%%%%%%%%%%%%%%%%%%
%%%%%%%%%%%%%%%%%%%%%%%%%%%%%%%%%%%%%%%%%%%%

Figure~\ref{fig:front-velocity-with-convection}~(b) shows the asymptotic front velocity over a large range of $\text{Ra}_s$. The filled symbols are results for fronts traveling through convection rolls. We do not include error bars here since the magnitude of the oscillations of $v_f(t)$ are on the order of the symbol size used in the figure. The open triangles are the results in the absence of thermal convection ($\text{Ra}_T\!=\!0$) and are included here for comparison. It is clear that for small and intermediate values of $\text{Ra}_s$, shown by the green diamonds and the one blue circle at $\text{Ra}_s\!=\!0$, that the front velocity remains constant in this regime.

However, for larger values of $\text{Ra}_s$, Fig.~\ref{fig:front-velocity-with-convection}~(b) shows that the front velocity increases and eventually is described by the $\text{Ra}_s^{1/2}$ scaling indicated by the dashed line. It is clear that in comparison with the front velocities in the absence of thermal convection (the open symbols in Fig.~\ref{fig:front-velocity-with-convection}~(b)) that the fronts with thermal convection have an increased velocity for all values of $\text{Ra}_s$. The increase in velocity is approximately constant where $\Delta \bar{v}_f \!=\! \bar{v}_f \!-\! \bar{v}_f(\text{Ra}_T\!=\!0) \!\approx\! 0.5$ for $\text{Ra}_s \gtrsim 2000$. 

Our findings indicate that propagating fronts with solutal feedback in the presence of counter-rotating thermal convection rolls  have a decreased mixing length, an increased front velocity,  an oscillating characteristic fluid velocity, and increased oscillations in the front velocity. These results are due to the complex interactions between the solutal feedback and the fluid dynamics.  The interactions between the front and the fluid can be further elucidated using space-time plots of the concentration field.

\begin{figure}[tbh]
   \begin{center}
   \includegraphics[width=3.3in]{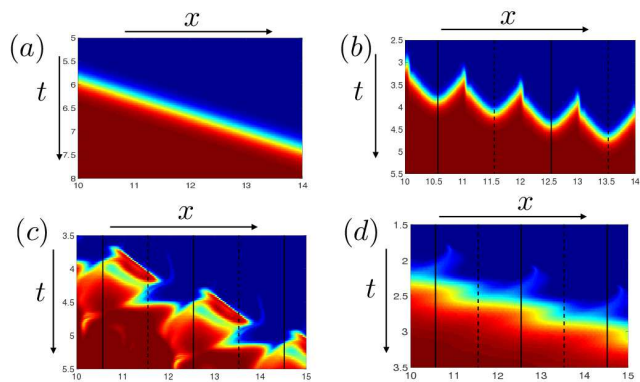} 
   \caption{The spatiotemporal features of propagating fronts. Space-time plots are shown of the concentration at the horizontal midplane $c(x,z\!=\!1/2,t)$ where $x$ is the horizontal axis and $t$ is the vertical axis. Red is products, blue is reactants, and the yellow/green regions indicate the reaction zone.  The spatial location of the thermal convection rolls are indicated by the vertical lines. The centers of convection rolls with a clockwise (counter-clockwise) rotation are shown with solid (dashed) lines. Only a small portion of space and time are shown in order to visualize the complex features. (a) Solutal feedback without thermal convection ($\text{Ra}_s\!=\!1000$, $\text{Ra}_T\!=\!0$). (b)~No solutal feedback with thermal convection ($\text{Ra}_s\!=\!0$, $\text{Ra}_T\!=\!3000$). Solutal feedback and thermal convection (c)~$\text{Ra}_s\!=\!1000$, $\text{Ra}_T\!=\!3000$; and (d)~$\text{Ra}_s\!=\!6000$, $\text{Ra}_T\!=\!3000$.}
    \label{fig:spacetime}
    \end{center}
\end{figure}

In Fig.~\ref{fig:spacetime} we show space-time plots of the concentration field at the horizontal midplane $c(x,z\!=\!1/2,t)$ where $x$ is the horizontal axis and $t$ is the vertical axis with positive time in the downward direction. Red is products, blue is reactants, and the reaction zone is the green/yellow region.  The vertical lines in Fig.~\ref{fig:spacetime}(b)-(d) indicate the locations of the centers of the convection rolls in the fluid before the front passes through where solid (dashed) indicates a clockwise (counter-clockwise) rotating convection roll.
%%%%%%%%%%%%%%%%%%%%%%%%%%%%%%%%%%%%%%%%%%%%
%%%%%%%%%%%%%%%%%%%%%%%%%%%%%%%%%%%%%%%%%%%%

A space-time plot for the case of $\text{Ra}_s\!=\!\text{Ra}_T\!=\!0$ (not shown) would simply yield a green/yellow region that is a line from the upper left to the lower right where the inverse slope of the line is the asymptotic front velocity $\bar{v}_f$. A similar result is obtained for $\text{Ra}_s \!>\! 0$ with $\text{Ra}_T\!=\!0$ as shown in Fig.~\ref{fig:spacetime}(a) for the specific case of $\text{Ra}_s\!=\!1000$ and $\text{Ra}_T\!=\!0$.  This linear picture changes significantly in the presence of thermal convection as shown in Fig.~\ref{fig:spacetime}(b)-(d).

The case with thermal convection, but without solutal feedback, is shown in Fig.~\ref{fig:spacetime}(b).  The space-time plot yields a periodic structure with triangular features. The troughs are located at the center of the convection rolls because the front spirals inward toward the roll centers which requires extra time. The peaks of the triangular structures occur at locations between convection rolls where the fluid velocity is either a maximum in the upward  or downward directions. For example, in Fig.~\ref{fig:spacetime}(b) a maximum downflow occurs at $x\!=\!11$ and a maximum upflow occurs at $x\!=\!12$.  In the absence of solutal feedback, the upflow and downflow regions yield symmetric triangular features in the spacetime plot. 

A horizontal slice through Fig.~\ref{fig:spacetime}(b) at any time $t$ would yield the spatial variation of the mid-plane concentration at that time. For example, one horizontal slice of Fig.~\ref{fig:spacetime}(b) corresponds to a mid-plane slice through the image shown in Fig.~\ref{fig:fronts-with-convection}(a) where it is clear that centers of the rolls are the last to complete the reaction and the convection roll edges are the first. A vertical slice through Fig.~\ref{fig:spacetime}(b) at any position $x$ would yield $c(t)$ at that location. It is clear that any vertical slice of Fig.~\ref{fig:spacetime}(b) would yield a monotonically increasing dependence for $c(t)$ as the reaction goes from reactants to products with increasing time at any particular location $x$.

This picture changes significantly in the presence of solutal feedback. Figure~\ref{fig:spacetime}(c) shows the spacetime plot for a front with both solutal feedback ($\text{Ra}_s\!=\!1000$) and thermal convection ($\text{Ra}_T\!=\!3000$). There are now considerable changes to the spatial and temporal variations of the concentration field. This front is also shown in Fig.~\ref{fig:fronts-with-convection}(e). An interesting feature is the emergence of temporal oscillations in the concentration field at particular locations. For example, a vertical slice at $x\!=\!11.5$ which corresponds with the vertical dashed line would yield a concentration that oscillates in time as it goes from reactants to products. There are also spatially complex regions in the product region where the reaction is slow to reach completion, for example near $x\!\approx\!12$ at time $t\!\approx\!5.5$.

Figure~\ref{fig:spacetime}(d) shows the space-time plot for a case where $\text{Ra}_s$ is large and the solutally driven flow dominates the convective flow. In this case, the space and time features are much smoother. However, small temporal oscillations of $c(t)$ are still present for particular choices of $x$ such as $x\approx 13$.  Although the front annihilates the convection rolls as it passes through, the leading edge of the front does interact directly with the convection rolls which leads to the wisp-like structures in light blue that indicate the locations where the reaction first takes place.  For example, a wisp is located near $x \! \approx \! 11$ and $t \! \approx \! 2$.  

\section{Conclusions}

We have used high-order numerical simulations to explore the dynamics of propagating fronts with solutal feedback for a range of conditions where the complex interactions between reaction, diffusion, and convection contributions are important. In the absence of an externally driven flow we quantified the solutally driven convection roll that propagates along with the front for a wide range of conditions. In the presence of counter-rotating convection rolls we investigated the interaction between this solutally driven convection roll and the thermal convection. 

In our study, we have used the incompressible Navier-Stokes equation for the fluid with the Boussinesq approximation to account for density changes due to thermal and solutal variations. The concentration field was described by a reaction-convection-diffusion equation with the addition of a FKPP nonlinearity.  Our approach is quite general and could be extended in a straightforward manner to include more complex features. For example, three-dimensional geometries, time-varying convective flow fields, large Reynolds numbers flows, and different forms of the nonlinear expression could be used to model the chemical reaction where many open questions remain.

However, a particularly interesting direction to explore is to a include a thermal contribution for the reaction. For example, a front propagating through an externally imposed flow field resulting from an exothermic autocatalytic reaction where the density of the products and reactants also vary. The dynamics resulting from these subtle interactions are expected to be quite rich and remain a topic of future interest. 

~\vspace{0.2cm}

\noindent Acknowledgments: We are grateful for many fruitful interactions with Paul Fischer. The numerical computations were done using the resources of the Advanced Research Computing center at Virginia Tech.

~\vspace{0.2cm}

\appendix*

\section{Numerical Approach used for Perturbation Analysis}
\label{section:numerics}

We briefly describe the numerical approach used to simulate the equations discussed in the perturbation analysis of Sec.~\ref{section:perturbation} for $\text{Ra}_s \!\ll\! 1$. The equations for $\psi$, $c$, and $\omega$ are numerically solved to $\mathcal{O}(2)$. We found that a fully-explicit finite-difference approach that is first order accurate in time and second order accurate in space was sufficient.

We numerically solve Eqs.~(\ref{eq:c0}),~(\ref{eq:omegapsi1})-(\ref{eq:cpsi2}) with the appropriate boundary and initial conditions described in Sec.~\ref{section:perturbation}. We use an equally spaced grid where $\Delta x \!=\! \Delta z \!=\! 0.02$ on a domain with an aspect ratio of $\Gamma \!=\! 12$. For time derivatives we use a first-order forward Euler time difference with a time step of $\Delta t \!=\! 1 \times 10^{-4}$. For spatial derivatives we used second order central time differencing.

The following procedure is used to evolve forward the variables for the concentration, stream function, and vorticity from time step $n$ to $n+1$ at each order of $\text{Ra}_s$. We evolve the equations in the sequence $\mathcal{O}(0)$, $\mathcal{O}(1)$, and then  $\mathcal{O}(2)$. It would be straightforward to continue at higher order if desired.

We first evolve forward Eq.~(\ref{eq:c0}) for the concentration to yield its value at the next time step $c_0^{(n+1)}$. We next solve Eq.~(\ref{eq:omegapsi1}) for the vorticity $\omega_1^{(n+1)}$ at all interior grid points. The stream function $\psi_1^{(n+1)}$ is then evaluated over the entire domain using Eq.~(\ref{eq:omega1}) and a Gauss-Seidel iterative solver. With $\psi_1^{(n+1)}$ computed, we then evaluate the vorticity $\omega_1^{(n+1)}$ at the boundaries using Thom's formula~\cite{thom:1933,weinan:1996}. The concentration $c_1^{(n+1)}$ is then evaluated using Eq.~(\ref{eq:cpsi1}).

A similar procedure is followed at $\mathcal{O}(2)$. The vorticity $\omega_2^{(n+1)}$ at all interior points is computed using Eq.~(\ref{eq:omegapsi2}) and $\psi_2^{(n+1)}$ is computed over the entire domain using the Poisson equation relating the stream function and vorticity at $\mathcal{O}(2)$. Finally, $\omega_2^{(n+1)}$ is computed at the boundaries using Thom's formula and $c_2^{(n+1)}$ is evaluated over the entire domain using Eq.~(\ref{eq:cpsi2}). The overall procedure is then repeated to integrate the concentration, stream function, and vorticity variables forward in time.

\end{document}